\documentclass[preprint,review,12pt]{elsarticle}

\usepackage{amssymb}
\usepackage{amsmath,bm}
\usepackage{colortbl,booktabs}
\usepackage{tabularx}
\usepackage{threeparttable}
\usepackage{multicol,multirow}
\usepackage{subfigure}
\usepackage [font=footnotesize,labelfont=bf]{caption}
\usepackage{rotating}
\usepackage{tikz}
\usepackage{hyperref}

\usepackage{color}

\journal{Elsevier}

\begin{document}

\begin{frontmatter}

	\title{An efficient implementation of the bidirectional buffer: towards laminar and turbulent open-boundary flows }

	\author{Feng Wang\texorpdfstring{}}
	\ead{feng.wang.aer@tum.de}
	\author{Xiangyu Hu\texorpdfstring{\corref{mycorrespondingauthor}}{}}
	\cortext[mycorrespondingauthor]{Corresponding author.}
	\ead{xiangyu.hu@tum.de}
	\address{Department of Mechanical Engineering, Technical University of Munich\\
		85748 Garching, Germany}

	\begin{abstract}

		To effectively handle flows characterized by strong backflow and multiple open boundaries within particle-based frameworks, this study introduces three enhancements to improve the consistency, independence, and accuracy of the buffer-based open boundary condition in SPHinXsys.
		First, to improve the buffer consistency, the continuum hypothesis is introduced to prevent the excessive particle addition induced by strong backflow.
		Secondly, the independence of the bidirectional buffer is enhanced through region-constrained and independent labeling schemes, which effectively eliminate buffer interference and erroneous particle deletion in complex open-boundary flows.
		Thirdly, the original zeroth-order consistent pressure boundary condition is upgraded to first-order consistency by introducing a mirror boundary treatment for the correction matrix.
		The implementation is based on the rigorously validated weakly compressible smoothed particle hydrodynamics coupled with Reynolds-averaged Navier–Stokes (WCSPH–RANS) method, and both laminar and turbulent flow simulations are performed.
		Four test cases, including straight and U-shaped channel flows, a plane jet, and the flow in a 3D self-rotational micro-mixer, are conducted to comprehensively validate the proposed improvements.
		Among these cases, the turbulent plane jet is successfully simulated at a moderate resolution within a very compact computational domain involving strong backflow, a condition that is usually challenging for mesh-based methods.
		The three improvements require only minor modifications to the code framework, yet they yield significant performance gains.

	\end{abstract}

	\begin{keyword}
		Smoothed particle hydrodynamics  \sep open boundary flow \sep consistency \sep turbulence \sep plane jet
	\end{keyword}

\end{frontmatter}
%
%
\section{Introduction}
\label{sec1}
The Lagrangian particle-based methods, such as the smoothed particle hydrodynamics(SPH) method, have been applied to simulating the internal flows, including arterial flows in bioengineering\cite{lu2024gpu} and pipe/channel flows in fluid machinery\cite{wang2022simulation}.
Due to the mesh-free, Lagrangian characteristics, the particle-based methods may disclose new flow mechanism and break through existing bottlenecks for the complex internal flow problems, particularly in challenging scenarios such as fluid-structure interactions\cite{khayyer2018enhanced,zhang2021sphinxsys} and multiphase flows\cite{khayyer2013enhancement}.
However, to fully realize the potential of the particle-based method in simulating complex internal flows, it is crucial to develop a robust open boundary condition which can adapt to various simulation scenarios, such as the flows with strong backflow or multiple inlet/outlet boundaries.

On the one hand, addressing the strong backflow is challenging for both the mesh-based and mesh-less method, while the latter approach generally encounters greater difficulties.
This is because, different from the traditional mesh-based methods where the open boundary condition(OBC) can be directly designated on the inlet/outlet surfaces, the implementation of the OBC for particle-based methods suffers one extra technical difficulty, that is the particle addition/deletion.
Without an appropriate way to add or delete particles, the condition of the continuous flow may be broken\cite{liu2019numerical}, and the simulation consistency may deteriorate.

A common strategy to achieve this operation is to set up a buffer region\cite{shakibaeinia2010weakly, aristodemo2015sph} not only to add or delete particles when inflow or outflow occurs, but also to avoid the kernel truncation for the inner fluid particles.

Building upon this buffer-based strategy, to further enhance stability and consistency, several other techniques have been proposed, such as the segment-based mirror\cite{kunz2016inflow,monteleone2017inflow}, ghost node\cite{tafuni2018versatile, verbrugghe2019non}, incompressible correction\cite{shibata2015boundary}, cell-based\cite{shibata2011transparent}, average point\cite{kazemi2017sph,zhang2019mixed}, semi-analytical\cite{ferrand2017unsteady} schemes.
Among these schemes, the local-relabeling-based buffer technique\cite{han2018sph,zhang2023lagrangian} is not only easy to implement, but also has relatively high computational efficiency, since no additional computational nodes, points, mirror particles or segments are involved.
The latest version improves the flexibility of this technique to handle the bidirectional\cite{zhang2025dynamical} and arbitrary directional\cite{zhang2024generalized} flows.

Nevertheless, the local-relabeling-based buffer scheme faces challenges in maintaining consistency in the presence of backflow.
When there are particles moving back and forth across the relabeling boundary, additional particles will be unlimited generated, which violates the original design principle and leads to simulation crash.
Please note that the backflow in this context denotes transient particle motions occurring near the boundary within very short time spans.
This situation differs from that in Reference\cite{zhang2025dynamical}, where the flow initially proceeds in one direction for a period before reversing, as in the pulsatile channel flow case.
Additionally, the backflow in our case refers not only to physically-induced phenomena but also to numerically-induced artifacts.
That means under the influence of the initial pressure wave, the fluid particles tend to wander near the relabeling boundary during the start-up stage, leading to the unphysical particle addition, as well.
Although there are some remedies to avoid the adverse effect caused by the backflow, such as increasing numerical dissipation\cite{zhang2017weakly} or adding a damping zone\cite{sun2019extension} near the outlet, to the best of our knowledge, none of the existing studies address this problem for the local-relabeling-based buffer scheme essentially.

On the other hand, most existing studies about open boundary flows focus on systems with a single inlet-outlet configuration or unidirectional flow, whereas multiple inlet/outlet boundaries and opposing-direction flows remain largely unexplored.
Although increasing the number of inlets or outlets seems straightforward, simulation failures may occur when the buffer regions interfere with each other.
Besides, for the multi-in/outlet flows, it is also challenging to design a general and concise code, especially for the local-relabeling-based buffer scheme.

In this work, to address the two aforementioned issues, we enhance the consistency and independence of the local-relabeling-based buffer technique for handling strong backflow and systems with multiple inlets and outlets.
Specifically, first, to essentially avoid the unlimited particle addition caused by the backflow, a small fringe region is introduced to offset the relabeling boundary and the particle creation threshold.
This method efficiently resolves the issue while incurring no extra computational overhead.
Moreover, it is highly compatible with the overall code framework, as it requires modifying only a small part of the code.

Second, to enhance the independence of the buffer regions, and avoid the interference issue, the original aligned box region\cite{zhang2024generalized} is limited by involving a simple contain checking function, and each buffer region is assigned a unique identifier (ID).
Additionally, the consistency of the dynamic pressure boundary condition is improved from the zeroth-order to the first-order by introducing the correction matrix, and a mirror boundary condition is imposed for the correction matrix near the open boundary.

Last but not least, several challenging benchmark cases are simulated to evaluate the proposed improvements.
These include the laminar/turbulent straight channel flow, turbulent U-shaped channel flow, laminar/turbulent plane jets, and a 3D self-rotational micro-mixer.
For the turbulent simulations, the rigorously validated weakly compressible SPH method coupled with the Reynolds-Averaged Navier-Stokes equations (WCSPH-RANS) is employed\cite{wang2025weakly}.
Besides, among these numerical tests, the plane jet case is studied in detail as a representative example.
To eliminate potential boundary interference, no wall boundaries are used, and six buffer regions are introduced to ensure numerical stability.
Converged results are successfully obtained, for the first time to the best of our knowledge, for the WCSPH-RANS method.

The remainder of this manuscript is organized as follows.
Section \ref{section-wcsph-rans} introduces the preliminary works, including the governing equations, numerical discretization, and the original open boundary condition.
The improvements are described in Section \ref{sec-improvements}.
Numerical examples are tested and discussed in Section \ref{section-numerical-examples}, and the concluding remarks are given in Section \ref{section-conclusion}.
The computational code of this work is released in the open-source SPHinXsys repository at https://github.com/Xiangyu-Hu/SPHinXsys.

%
%
\section{The WCSPH method for both laminar and turbulent simulations}
\label{section-wcsph-rans}
\subsection{Governing equations}
The conservation equations of mass and momentum for incompressible laminar and turbulent flows\cite{wang2025weakly} in the Lagrangian framework are
\begin{equation}
	\frac{\text{d} \rho}{\text{d} t} =  -\rho \nabla\cdot \mathbf v,
	\label{mass-equ}
\end{equation}
\begin{equation}
	\frac{\text{d} \mathbf v}{\text{d} t} =  - \frac{1}{\rho}\nabla p_{eff} +\nabla\cdot (\nu_{eff}\nabla \mathbf v) + \mathbf g,
	\label{momentum-equ}
\end{equation}
where $\mathbf{v}$ is the velocity, $\rho$ is the density,
$\frac{\text d}{\text d t}=\frac{\partial}{\partial t} + \mathbf v \cdot \nabla$ stands for material derivative.
$p_{eff}$ and $\nu_{eff}$ are the effective pressure and kinematic viscosity, respectively, and the expressions of the two variable differ in the laminar and turbulent simulations, as concluded in Table \ref{tab-unified-p-nu}, where $k$ refers to the turbulent kinetic energy, $\nu_l$ and $\nu_t$ are the kinematic molecular and eddy viscosity.

\begin{table}
	\centering
	\footnotesize
	\caption{Unified expressions for the effective pressure and kinematic viscosity.}
	\renewcommand{\arraystretch}{1.2}
	\begin{tabular}{l c c}
		\hline
		Viscous model    & $p_{eff}$                          & $\nu_{eff}$                 \\
		\hline
		Laminar          & $p$                                & $\nu_l$                     \\
		Turbulent (RANS) & $p_{eff} = p + \frac{2}{3} \rho k$ & $\nu_{eff} = \nu_l + \nu_t$ \\
		\hline
	\end{tabular}
	\label{tab-unified-p-nu}
\end{table}

To ensure incompressibility, a stiff isothermal equation of state is used, as
\begin{equation}
	p = \rho_0 c^2_0 \left( \frac{\rho}{\rho_0} - 1\right),
	\label{eos}
\end{equation}
where $\rho_0$ is the reference density and $c_0$ refers to the sound speed.

For turbulent simulation only, the two-equation $k - \epsilon$ RANS model is adopted, and the additional transport equations are
\begin{equation}
	\frac{\text{d} k}{\text{d} t} = \bm{\tau_t}\nabla \mathbf{v} -\epsilon+\nabla\cdot (D_k\nabla k),
	\label{k-equ}
\end{equation}
\begin{equation}
	\frac{\text{d} \epsilon}{\text{d} t} = C_1 \frac{\epsilon}{k} \bm{\tau_t}\nabla \mathbf{v} -C_2 \frac{\epsilon^2}{k}+\nabla\cdot (D_\epsilon\nabla \epsilon),
	\label{epsilon-equ}
\end{equation}
where $D_k = \nu_l+\nu_t/\sigma_k$ and $D_\epsilon = \nu_l+\nu_t/\sigma_\epsilon$ are the diffusion coefficients for $k$ and $\epsilon$, respectively.
$\bm{\tau_t} =\nu_t(\nabla\mathbf v+\nabla\mathbf v^T)  - 2 k \mathbf I/3$ is the Reynolds stress tensor.
The kinematic eddy viscosity is calculated by $\nu_t= C_{\mu} k^2 / \epsilon $.
The empirical constants including $C_1$, $C_2$, $C_{\mu}$, $\sigma_k$ and $\sigma_\epsilon$ are the same from the original $k-\epsilon$ version\cite{launder1983numerical}, and are listed in Table \ref{tab-coeff-ke} in the appendix.

\subsection{Numerical discretization}
To increase stability, the continuity equation is discretized based on a low-dissipative Riemann solver\cite{zhang2017weakly}, as expressed by
\begin{equation}
	\frac{\text{d} \rho_i}{\text{d} t} = 2 \rho_i \sum_{j}  (\mathbf{v_i}-\mathbf{v^*}) \nabla W_{ij} V_j,
	\label{discretize-continuity-equ-riemann}
\end{equation}

Here, $\mathbf{v^*}= U^* \mathbf{e}_{ij}+(\overline{\mathbf{v}}_{ij}-\overline{U}_{ij}\mathbf{e}_{ij})$, and the gradient of the kernel function is expressed as $\nabla W_{ij} = \frac{\partial W_{ij}}{\partial r_{ij}} \mathbf{e}_{ij}$, where $W_{ij}$ represents $W(\mathbf{r}_{ij}, h)$ and $h$ is the smoothing length that is fixed at 1.3$dp$, and $dp$ is particle spacing.
$\overline{U}_{ij}$ is the projection of the inter-average velocity $\overline{\mathbf{v}}_{ij}$ along the pairwise direction.

Note that the following three operator notations are used consistently throughout this work:
$(\phi)_{ij} = [(\phi)_i - (\phi)_j]$ refers to the pairwise difference;
$\overline{(\phi)}_{ij} = [(\phi)_i + (\phi)_j]/2$ means the particle-pair inter average;
$\widetilde{(\phi)}_{ij} = 2(\phi)_i  (\phi)_j/[(\phi)_i + (\phi)_j]$ refers to the pairwise harmonic average.

The intermediate velocity $U^*$ is calculated by
\begin{equation}
	U^*  = \overline{U}_{ij} + \frac{P_{ij}}{2\rho_0 c_0},
	\label{eq-intermediate-vel}
\end{equation}
where the subscript $_0$ means the reference value.
The discretization of the momentum equation mainly involves two terms: the acceleration induced by the pressure gradient and viscosity, and the general discretized form of Equation \eqref{momentum-equ} is

\begin{equation}
	\frac{\text{d} \mathbf v_i}{\text{d} t} =
	\left(\frac{\text{d} \mathbf v_i}{\text{d} t}\right)^p
	+\left(\frac{\text{d} \mathbf v_i}{\text{d} t}\right)^\nu
	+\mathbf{g}.
	\label{eq-discretize-momentum-general}
\end{equation}

For the pressure gradient induced acceleration, the reverse kernel gradient correction\cite{zhang2025towards} is adopted to ensure consistency, as expressed by
\begin{equation}
	\left(\frac{\text{d} \mathbf v_i}{\text{d} t}\right)^p = -\sum_{j} m_j
	\left(
	\frac{p_i \mathbf{B}_j+p_j \mathbf{B}_i}{\rho_i\rho_j}
	\right)	\cdot \nabla W_{ij},,
	\label{eq-discretize-momentum-p-RKGC}
\end{equation}
where $\mathbf{B}$ is the correction matrix calculated by
\begin{equation}
	\mathbf{B}_i = \left( - \sum_j \mathbf{r}_{ij} \otimes \nabla W_{ij} V_j \right)^{-1}.
	\label{equ-B-matrix}
\end{equation}

For the viscosity induced acceleration, the adaptive Riemann-eddy dissipation (ARD) scheme\cite{wang2025weakly} is used, as expressed by
\begin{equation}
	\left(\frac{\text{d} \mathbf v_i}{\text{d} t}\right)^{\nu}=
	\frac{2}{\rho_i}\sum_{j} \mu_{ad} \frac{\mathbf{v}_{ij}}{r_{ij}}\frac{\partial W_{ij}}{\partial r_{ij}}V_j,
	\label{eq-discretize-momentum-ARD2}
\end{equation}
where $\mu_{ad}$ is the adaptive viscosity that is computed by
\begin{equation}
	\mu_{ad}= max(\widetilde{\mu}_{ij}, \mu_{R}),
	\label{eq-discretize-momentum-ARD-viscosity}
\end{equation}

Here, $\mu_{R} = \frac{1}{2}\beta_{ij}\rho_0 c_0 h$ is the numerical viscosity, and $\beta_{ij} = \min( \eta \max(\mathbf{v}_{ij} \cdot\mathbf{e}_{ij},0), c_0 )$ is the dissipation limiter.
$\eta = 3$ is an empirical parameter that originates from Ref. \cite{zhang2017weakly}, and this value is determined according to the numerical tests, and used throughout this work.
Please note that for laminar simulations, the $\beta_{ij}$ is set as 0, and hence Equation \eqref{eq-discretize-momentum-ARD2} reduces to the origial pariwise viscous formulation as reported in Reference \cite{hu2006multi}.

The discretization of the $k$ and $\epsilon$ transport equations involves the approximation of the velocity gradient and the diffusion terms.
The velocity gradient is discretized by
\begin{equation}
	\nabla \mathbf{v}_i=\sum_{j} \mathbf{v}_{ij} \otimes (\mathbf{B}_i \nabla W_{ij}) V_j.
	\label{velo-grad-equ}
\end{equation}

The discretization of the diffusion terms in the $k$ and $\epsilon$ equations is analogous to that of the viscous term in the momentum equation.
Consequently, the discretized formulations of the two transport equations are written as
\begin{equation}
	\frac{\text{d}  k_i}{\text{d} t}=G_k -\epsilon_i
	+\frac{2}{\rho_i}\sum_{j}^{N} \widetilde{(D_{k})}_{ij}  \frac{k_{ij}}{r_{ij}}V_j\frac{\partial W_{ij}}{\partial r_{ij}},
	\label{k-discretized-equ}
\end{equation}

\begin{equation}
	\frac{\text{d} \epsilon_i}{\text{d} t} = C_1 \frac{\epsilon_i}{k_i} G_k -C_2 \frac{\epsilon_i^2}{k_i}
	+\frac{2}{\rho_i}\sum_{j}^{N} \widetilde{(D_{\epsilon})}_{ij} \frac{\epsilon_{ij}}{r_{ij}}V_j\frac{\partial W_{ij}}{\partial r_{ij}},
	\label{epsilon-discretized-equ}
\end{equation}
where $G_k$ is the discretized generation term of the turbulent kinetic energy.

\subsection{The wall boundary conditions and schemes for stability and efficiency}
The wall boundary condition of the pressure gradient term is based on the Riemann solver\cite{zhang2017weakly}.
As for the boundary condition of the viscous term, the non-slip condition is enforced for the laminar simulation, while for the turbulent simulation, the step-wise wall function method is adopted.
The details of the Lagrangian particle-based wall function implementation can be found in Ref. \cite{wang2025weakly}.

The transport velocity formulation\cite{zhang2017generalized} is used to avoid the tensile instability, and the duel-criteria time stepping scheme\cite{zhang2020dual} is adopted to increase computational efficiency.
\subsection{The local-relabeling-based open boundary condition}
\label{sec-pre-obc}
This section not only introduces the principle of the open boundary condition in SPHinXsys, but also presents its memory management optimization which contributes to improved computational efficiency.

\subsubsection{Configuration of the buffer region}
\label{sec-pre-define-buffer}
The determination of the buffer regions is based on the local cell link list\cite{zhang2024generalized} and the relabeling boundary, as shown in Fig. \ref{fig-preliminary-inlet-outlet-idea} (a).
Before the simulation starts, the buffer shape is defined by the user, and the corresponding local cell link list (CLL) is extracted from the global CLL.
Once the local CLL is defined, each operation involving the open boundary condition will merely search the particle information stored in the local CLL, which significantly reduce the computational amount, since checking all the particles is time-consuming.

Consequently, the buffer region is defined based on the local CLL and the two relabeling boundaries.
Particles located within this region are continuously relabeled as buffer particles, while those outside remain unaffected.
This locality is the reason why the scheme is referred to as local relabeling.

\subsubsection{Inflow/particle-addition}
\label{sec-pre-inflow}

\begin{figure}[htb!]
	\centering
	\includegraphics[trim = 11.77cm 0cm 0cm 0cm, clip,width=1.0\textwidth]{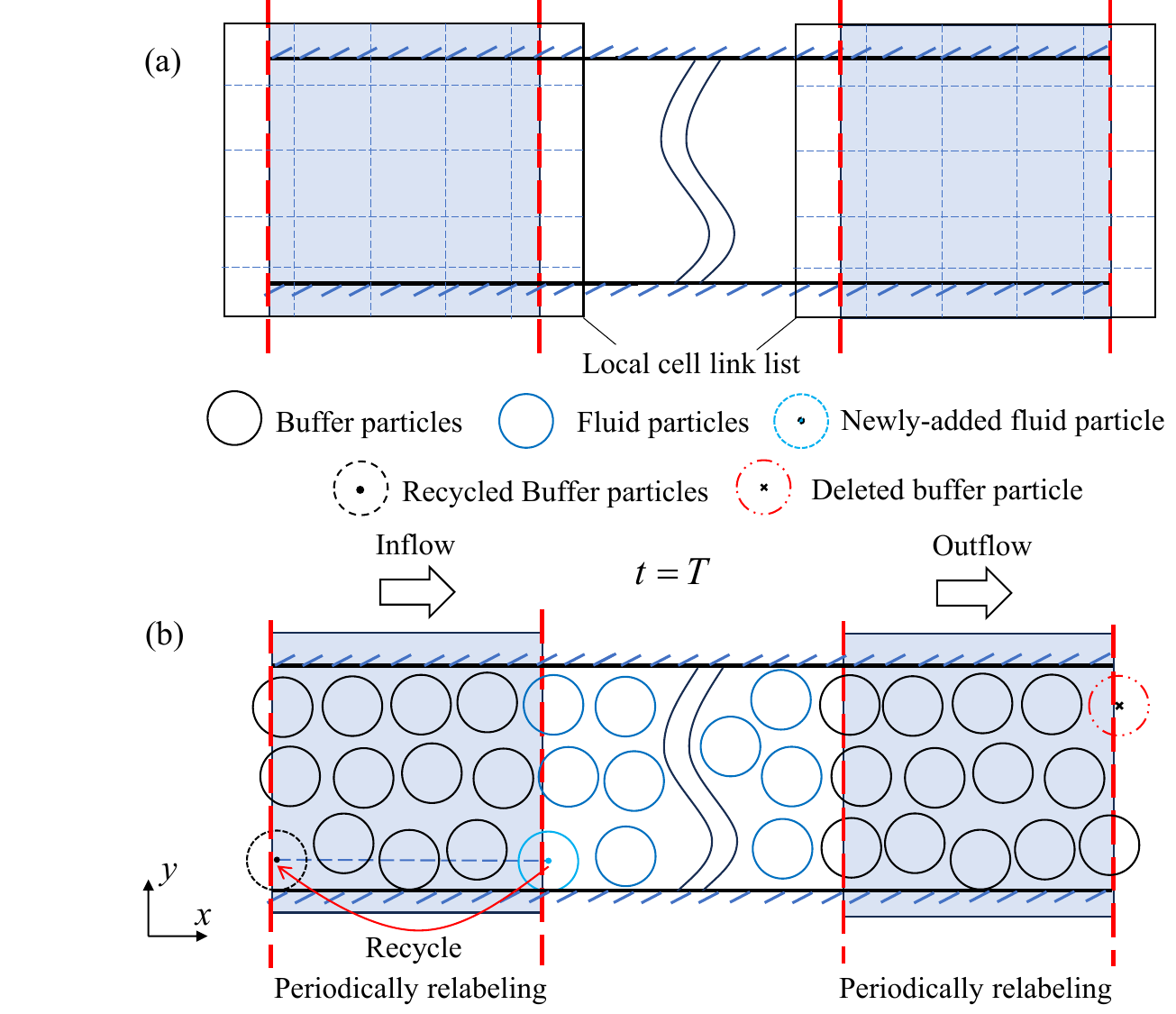}
	\caption
	{
		(a) The determination of the buffer regions by cell link list.
		(b) The concept of the particle treatment of the local-relabeling-based open boundary strategy at a specific time instant $T$ when both the inflow and outflow occur.
		This illustration is based on a straight channel flow.
		The red dot lines refer to the relabeling boundaries, and the blue regions are the buffer regions.
	}
	\label{fig-preliminary-inlet-outlet-idea}
\end{figure}

As for inflow, once a buffer particle crosses the relabeling boundary, a new fluid particle will be generated based on the moved-out buffer particle, as shown in Fig. \ref{fig-preliminary-inlet-outlet-idea} (b).
In summary, the pre-conditions for generating particles are:
(1) a particle crosses the relabeling boundary;
(2) its identity is "buffer".

The position and material properties of the newly-added fluid particle are inherited from those of the corresponding buffer particle that has crossed the boundary.
After the generation, the buffer particle will be recycled, re-entering the buffer region from the opposite end.
The boundary values of the recycled buffer particle are given based on the specified boundary conditions, such as velocity or pressure inlets. \cite{zhang2023lagrangian,zhang2025dynamical}, and its position is determined by
\begin{equation}
	\mathbf{r}_{rec}=\mathbf{r}_{in}-L_b \mathbf{n}_b,
	\label{eq-pos-new-added}
\end{equation}
where $\mathbf{r}_{rec}$ is the position of the recycled buffer particle and $\mathbf{r}_{in}$ is the fluid particle that was just created, $L_b$ and $n_b$ refer to the length and unit flow direction vector of this buffer, respectively.

It is worth noting that the information of the newly-added fluid particle is consistently stored at the end of the memory block in the code implementation.
Consequently, the storage indices of the existing particles remain unchanged, ensuring data consistency and minimizing memory reallocation overhead.

\subsubsection{Outflow/particle-deletion}
\label{sec-pre-outflow}
For outflow treatment, removing the outflow particle and clearing its associated data are straightforward; however, maintaining a continuous memory layout after deletion without triggering data reallocation can be challenging.
In SPHinXsys, an efficient particle deletion scheme based on data overwriting is adopted,

\begin{figure}[htb!]
	\centering
	\includegraphics[trim = 11.15cm 0cm 0cm 0.63cm, clip,width=0.7\textwidth]{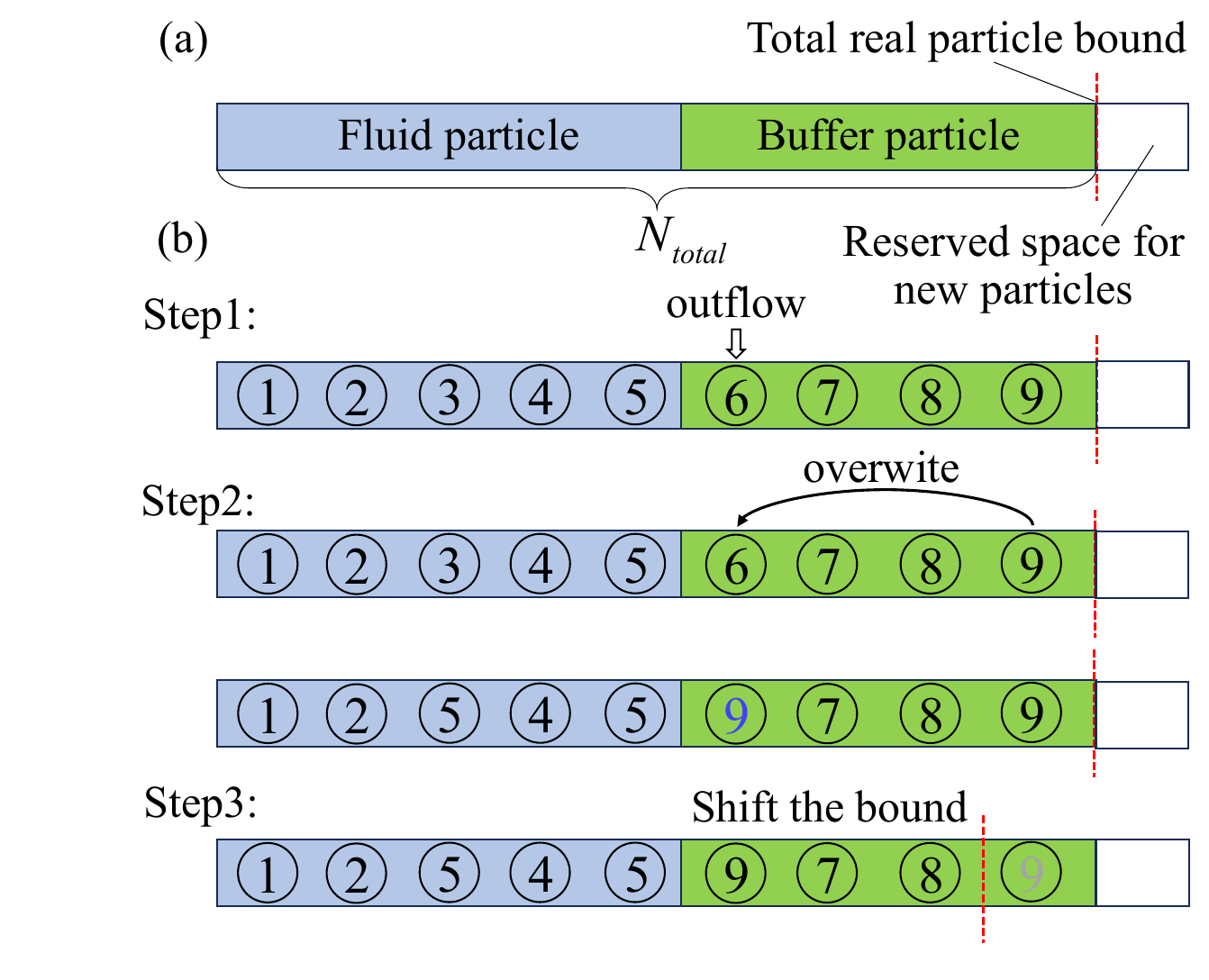}
	\caption
	{
		(a) Configuration of a whole particle array, and
		(b) the outflow steps when the particle 6 outflows.
	}
	\label{fig-outlet-delete-memory}
\end{figure}

The particle array configuration used in the flow simulation is illustrated in Fig. \ref{fig-outlet-delete-memory} (a).
Each array contains two types of particles, fluid and buffer particles, along with a pre-allocated space reserved for newly added particles.
The fluid and buffer particles are collectively referred to as real particles, with their total number denoted by $N_{total}$.
The outflow steps are demonstrated in Fig. \ref{fig-outlet-delete-memory} (b), when an arbitrary buffer particle, namely particle 6, outflows.

Step 1 is to determine whether a particle has moved out of the domain, typically using a local position check \cite{zhang2024generalized}.
Step 2 involves copying the information of the last fluid particle to the outflow particle, i.e., overwriting the data of particle 6 with that of particle 9.
The last step is to update the total number of the real particles by shifting the particle boundary forward accordingly.
These three steps can be implemented using a simple \verb|while| loop, and any modification to the memory order is avoided.
Please note that the particle index disorder does not affect simulation because the interaction between particles is based on the cell link list which is updated after the deletion at each advection time step \cite{zhang2021sphinxsys}.
\section{Improvements for complex open boundary flows}
\label{sec-improvements}
\subsection{Improvement on buffer consistency based on continuum hypothesis}
\begin{figure}[htb!]
	\centering
	\includegraphics[trim = 10.35cm 0cm 0cm 2.21cm, clip,width=1.0\textwidth]{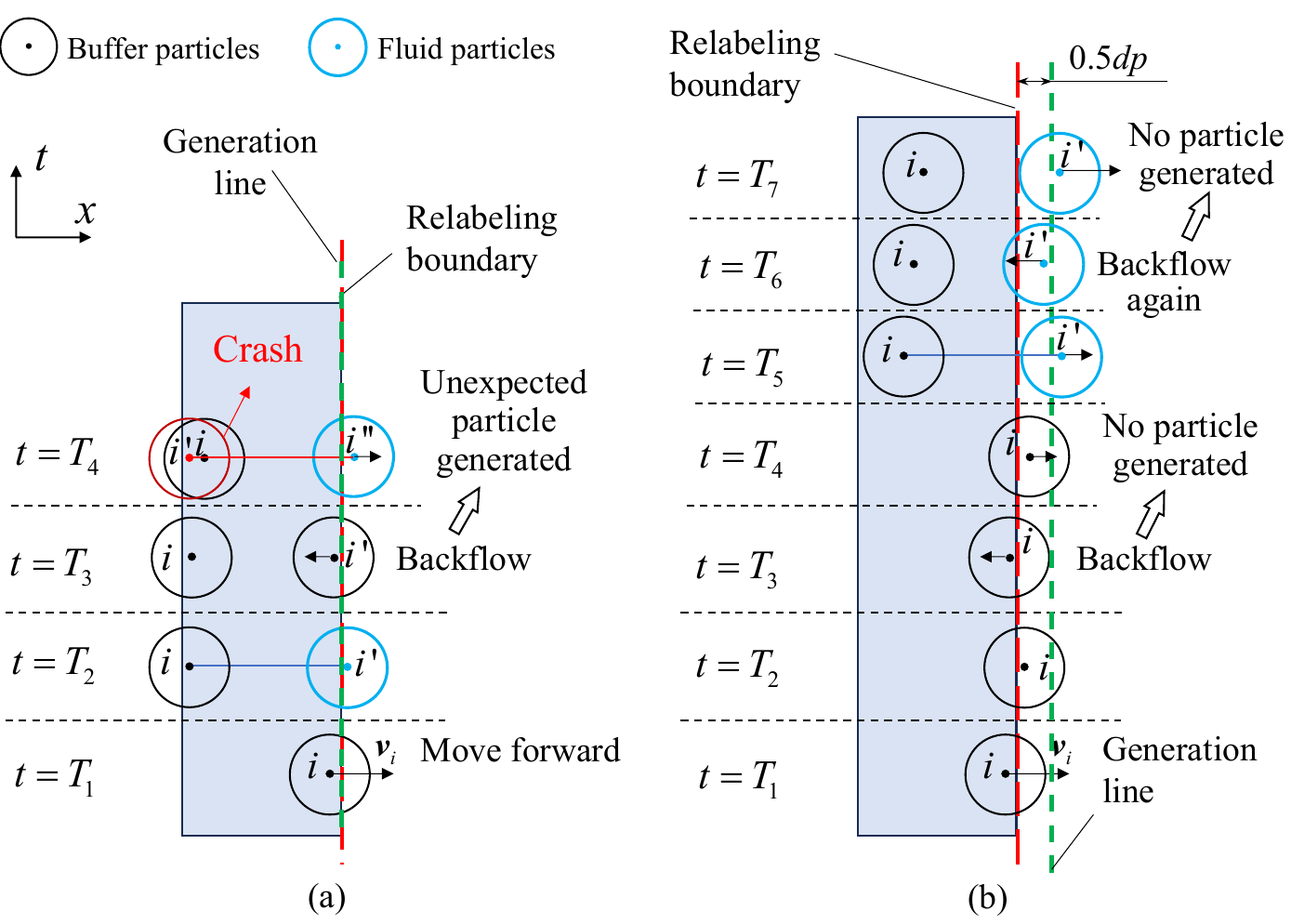}
	\caption
	{
		The continuous trajectory of a buffer particle $i$ located near the relabeling boundary and influenced by the backflow:
		(a) without the proposed improvement;
		(b) with the proposed improvement.
	}
	\label{fig-improve-buffer-consistency}
\end{figure}

The local-relabeling-based bidirectional buffer, as demonstrated in Section \ref{sec-pre-obc}, performs well for unidirectional and reversed unidirectional flows \cite{zhang2025dynamical}, but may encounter difficulties when backflow occurs.
As shown in Fig. \ref{fig-improve-buffer-consistency} (a), the continuous 4 time instants are observed and the motion of an arbitrary buffer particle $i$ near the relabeling boundary is tracked.
Please note that the empty space is also filled with particles which are not shown to better follow the movement of the particle $i$.

Initially, at the time instant $\text{T}_1$, the buffer particle $i$ moves forward, and then crosses the relabeling boundary at $\text{T}_2$.
Since the two preconditions for adding fluid particles mentioned in Sec. \ref{sec-pre-inflow} are satisfied, a new fluid particle $i'$ is generated at the position of $i$, and after the generation, the particle $i$ is moved to the left side of the buffer region.
Subsequently, at $\text{T}_3$, the particle $i'$ moves back due to the backflow and its identity becomes "buffer" again due to the relabeling.
Then at $\text{T}_4$, because of the complicated flow condition, before the particle $i$ is squeezed out of the buffer region, the buffer particle $i'$ immediately moves forward and crosses the relabeling boundary, which triggers the particle generation again.
Consequently, the buffer particle $i'$ is recycled back again, and a new fluid particle $i''$ is generated.
Even worse, the two buffer particles, $i$ and $i'$, overlaps, leading to the simulation crash.

The root cause of this problem, we believe, is that the continuum assumption of fluid flow is neglected.
In fact, each SPH particle represents a finite volume of fluid, with its influence domain defined by the kernel truncation radius.
Therefore, treating an SPH particle as a material point and immediately switching its identity when it crosses the relabeling boundary violate the continuum assumption, breaking the simulation consistency.

To address this problem and restore the consistency, we propose a simple but effective boundary-shifting scheme, as shown in Fig. \ref{fig-improve-buffer-consistency} (b).
The general idea is to maintain the particle identity for a short time, after the particle crosses the boundary.
Specifically, the relabeling boundary (line) is no longer regarded as the criteria for generating particles, instead, the particle generation line is introduced.
The two lines, which were originally regarded as aligned, are now intentionally staggered by a small distance, namely $0.5dp$.
That means the criteria for particle generation are updated as follows:
(1) the particle crosses the particle generation line;
(2) its identity is "buffer".

To clearly present the effect of the improvement, similar to that demonstrated in Fig. \ref{fig-improve-buffer-consistency} (a), we also provide the movement of the particle $i$ after applying the consistency improvement, as shown in Fig. \ref{fig-improve-buffer-consistency} (b).
The trajectory of the particle $i$ during the time instants $\text{T}_1 - \text{T}_4$ is the same as that shown in Fig. \ref{fig-improve-buffer-consistency} (a), while no particles are generated because the generation precondition (1) is not satisfied.
Until $\text{T}_5$, when the buffer particle $i$ crosses the generation line, a new fluid particle $i'$ is created and the particle $i$ is recycled.
Subsequently, at $\text{T}_6$, backflow happens again, and the fluid particle $i'$ moves back and crosses the generation line from the opposite direction.
However, this time the identity of the particle $i'$ is not changed and is still "fluid" since the relabeling merely occurs in the buffer region bounded by the relabeling boundary.
Therefore, at $\text{T}_7$, no particle will be generated although the fluid particle $i'$ re-crosses the generation boundary, because the precondition (2) is not satisfied.
As a result, the unexpected particle generation problem is well addressed.

Please note that the small offset distance, 0.5$dp$, accounts for the continuous property and is determined by numerical tests.
This value is consistently used throughout this study and is generally effective for all the open boundary flow cases in SPHinXsys library.
Besides, as shown in Fig. \ref{fig-improve-buffer-consistency}(b) at $\text{T}_6$, if the backflow is strong enough that the fluid particle $i'$ crosses the relabeling boundary from the opposite direction, the offset distance should be sufficiently large to ensure that the leftmost buffer particle $i$ has already been squeezed out of the buffer region and deleted.

\subsection{Improvement on buffer independence for multiple in/outlets}
\label{sec-independence}
For systems with multiple inlets and outlets, the original local-relabeling-based buffer may encounter two issues caused by buffer interference.
This section discusses the issues and provides simple yet effective solutions to improve the independence of each buffer.

\begin{figure}[htb!]
	\centering
	\includegraphics[trim = 12.72cm 0cm 0cm 6.14cm, clip,width=0.9\textwidth]{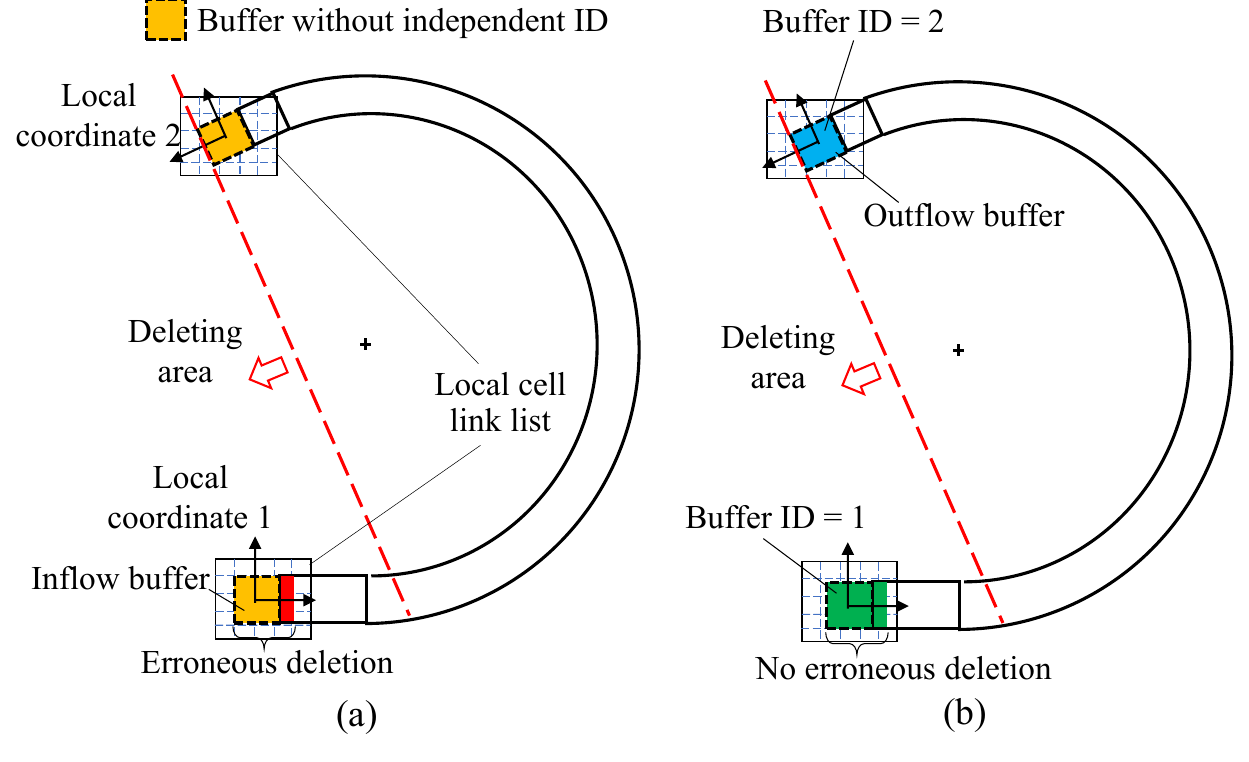}
	\caption
	{
		(a) Without unique buffer IDs, particle deletion errors may occur due to interference;
		(b) with unique buffer IDs, erroneous deletion is avoided.
	}
	\label{fig-independence-curved_channel_idea}
\end{figure}

The first problem is the erroneous deletion of the buffer particles in other buffer regions.
As demonstrated in Fig. \ref{fig-independence-curved_channel_idea}, for this strongly-curved channel, the bending angle is more than 180$^\circ$, making the inflow buffer exposed in the deleting area of the outflow buffer, since the deletion area is determined by the relabeling boundary after coordinate transfer.
Furthermore, as shown in Sec. \ref{sec-pre-define-buffer}, because the local cell link lists are originated the same global CLL, the outflow check is performed not only for particles within the outflow buffer but also for those in the inflow buffer.
Consequently, particles in or near the inflow buffer are continuously deleted, eventually leading to a simulation crash.
Additionally, the newly added particles are erroneously deleted first, as they are placed at the end of the memory array.

\begin{figure}[htb!]
	\centering
	\includegraphics[trim = 16.23cm 0cm 0cm 7.09cm, clip,width=0.8\textwidth]{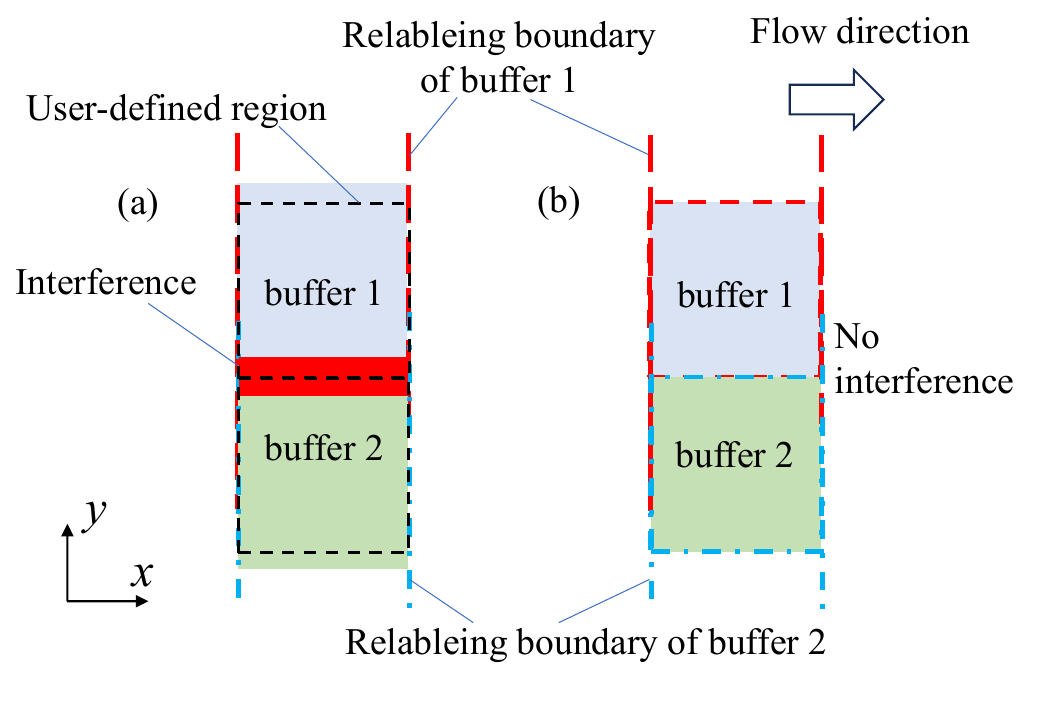}
	\caption
	{
		When the two buffer regions are vertically stacked,
		(a) interference occurs along $y$ direction;
		(b) interference is avoided after adding the contain-checking.
	}
	\label{fig-independence-stack-buffers}
\end{figure}

The second issue is the overlap between the two buffer regions when they are placed adjacently, even though the user-defined regions themselves do not overlap.
As shown in Fig. \ref{fig-independence-stack-buffers} (a), when the two buffers are vertically stacked, their effective areas will overlap and interfere with each other along the vertical direction.
This is because the effective area of each buffer is determined by the cell-linked list (CLL), and each cell in the CLL is generally larger than the particle spacing \cite{zhang2021sphinxsys}.
Therefore, the area defined by the cell link list not only covers the user-defined buffer region but also slightly extends beyond it.
Although the buffer size along the main flow direction is constrained by the relabeling boundary, it remains unrestricted in the direction perpendicular to the main flow direction \cite{zhang2024generalized}.
An additional example can be seen in Fig. \ref{fig-preliminary-inlet-outlet-idea}(a), where the buffer region extends beyond the wall, contrary to expectations.
Consequently, placing the buffer along an unrestricted direction, such as the vertical direction illustrated in Fig. \ref{fig-independence-stack-buffers} (a), may lead to interference.

To address the issue of the erroneous deletion, a unique identifier, namely the buffer ID, is assigned to each buffer at the configuration stage, and the buffer particles are tagged with the corresponding identifier during relabeling, as shown in Fig. \ref{fig-independence-curved_channel_idea} (b).
Furthermore, an additional check between the buffer ID and particle ID is incorporated into the first step of particle deletion that is described in Sec. \ref{sec-pre-outflow}.
A particle will be deleted only if its particle ID matches the buffer ID.
With this scheme, the unexpected deletion is effectively avoided, while introducing almost no additional computational cost and requiring only minor modifications to the code framework.

To prevent interference between effective domains, a containment checking function is introduced to restrict each spatial direction of the user-defined buffer region, as shown in Fig. \ref{fig-independence-stack-buffers} (b).
Given that the buffer region is predefined as a rectangle (in 2D) or a cuboid (in 3D), the containment test can be efficiently implemented by applying a coordinate transformation\cite{zhang2024generalized} and performing a direct box-inclusion query.
Specifically, the box-inclusion check is performed by testing whether the transformed point lies within a reference box using a built-in geometric method for axis-aligned box containment testing.

With the two improvements on buffer independence, the local-relabeling-based open boundary condition becomes capable of handling complex flow problems with more intricate inflow and outflow configurations.
Please note that although the demonstrations are two-dimensional, the proposed improvements are applicable to both 2D and 3D simulations.

\subsection{Improvement on the accuracy of the open boundary flow by introducing the RKGC with a mirror boundary condition}
\label{sec-accuracy-improvement}
The RKGC scheme was firstly proposed for reducing the unphysical numerical dissipation in free surface flows without the open boundary \cite{zhang2024towards}.
Later, the importance and necessity of this technique in simulating both the gravity-driven free-surface and pressure-driven channel flows have been proved \cite{wang2025residue}.
However, when applying this technique to open boundary flows, the boundary condition for the correction matrix $\mathbf{B}$ requires careful treatment; otherwise, unexpected results may occur.
Given that RKGC primarily affects the pressure gradient approximation in fluid dynamics, this section introduces its near-boundary treatment, with the mirror boundary condition serving as the underlying strategy.

As shown in Equation \eqref{eq-discretize-momentum-p-RKGC}, the pressure gradient is approximated by combining the correction matrix.
For a fluid particle $b$ located near the open boundary, if not consider the wall, the pressure gradient consists of two components: the contribution from the internal neighboring particles and the compensation term imposed by the pressure boundary condition \cite{zhang2025dynamical}
\begin{equation}
	\nabla p_{b}  =
	\sum_{j=1}^{N_{inter}}\frac{m_{j}}{\rho_{j}}(p_i \mathbf{B}_j+p_j \mathbf{B}_i) \cdot \nabla W_{ij} +
	\sum_{k=1}^{N_{ob}} \frac{m_{k}}{\rho_{k}} (p_i \mathbf{B}_k+p_k \mathbf{B}_i)  \cdot \nabla W_{ik}.
	\label{eq-pressure-gradient-obc-B}
\end{equation}
where subscript $_{inter}$ and $_{ob}$ refer the internal neighbor fluid particles and the open boundary (ghost) particles, respectively.
Please note that the pressure boundary condition discussed herein refers not only to the prescribed pressure inlet or outlet conditions, but also to the commonly adopted extrapolation treatment.

To calculate the modified compensation term, the mirror boundary on the correction matrix, $\mathbf{B}_k=\mathbf{B}_i$, is imposed, and hence the second term on the right-hand side of Equation \eqref{eq-pressure-gradient-obc-B} is modified as
\begin{equation}
	\sum_{k=1}^{N_{ob}} \frac{m_{k}}{\rho_{k}} (p_i +p_k ) \mathbf{B}_i \cdot \nabla W_{ik}.
	\label{eq-pressure-gradient-RHS2-obc-B}
\end{equation}

The pressure at the open boundary, $p_{ob}$, is imposed on each particle pair, such that $\overline{p}_{ik} = p_{ob}$.
It should be noted that $p_{ob}$ can be manually designated or extrapolate from the fluid domain.
To compute Eq. \eqref{eq-pressure-gradient-RHS2-obc-B}, the originally-used zero-order consistency condition is improved as
\begin{equation}
	\sum_{j=1}^{N_{inter}}\frac{m_{j}}{\rho_{j}} \mathbf{B}_i \cdot \nabla W_{ij}+
	\sum_{k=1}^{N_{ob}}\frac{m_{k}}{\rho_{k}} \mathbf{B}_i \cdot \nabla W_{ik}
	\approx  \mathbf{0},
	\label{eq-first-order-consistency}
\end{equation}

Substituting the open boundary pressure and Eq. \eqref{eq-first-order-consistency} into Eq. \eqref{eq-pressure-gradient-obc-B}, we have
\begin{equation}
	\nabla p_{b}  =
	\sum_{j=1}^{N_{inter}}\frac{m_{j}}{\rho_{j}}(p_i \mathbf{B}_j+p_j \mathbf{B}_i) \cdot \nabla W_{ij} -
	2 p_{ob}\sum_{j=1}^{N_{ob}} \frac{m_{j}}{\rho_{j}} \mathbf{B}_i  \cdot \nabla W_{ij}.
	\label{eq-pressure-gradient-obc-B-final}
\end{equation}
%

\section{Numeral examples}
\label{section-numerical-examples}
\subsection{Fully developed flow in a straight channel}
The fully developed straight channel flow is a classical benchmark case for verifying the stability and accuracy of the open-boundary treatments.
This section will test both the laminar and turbulent flows in the straight channel, and demonstrate the effectivity of the improvement mentioned in Sec. \ref{sec-accuracy-improvement}.
The Reynolds number is defined using the channel width and the bulk velocity.
The number of fluid particles across the cross-section is 20.

As for the laminar simulation, the Reynolds number is 50.
The parabolic velocity inlet condition is imposed, in which the maximum velocity is 0.0125.
The outlet pressure is 0.1.

\begin{figure}[htb!]
	\centering
	\includegraphics[trim = 6.01cm 0cm 0cm 5.51cm, clip,width=1.0\textwidth]{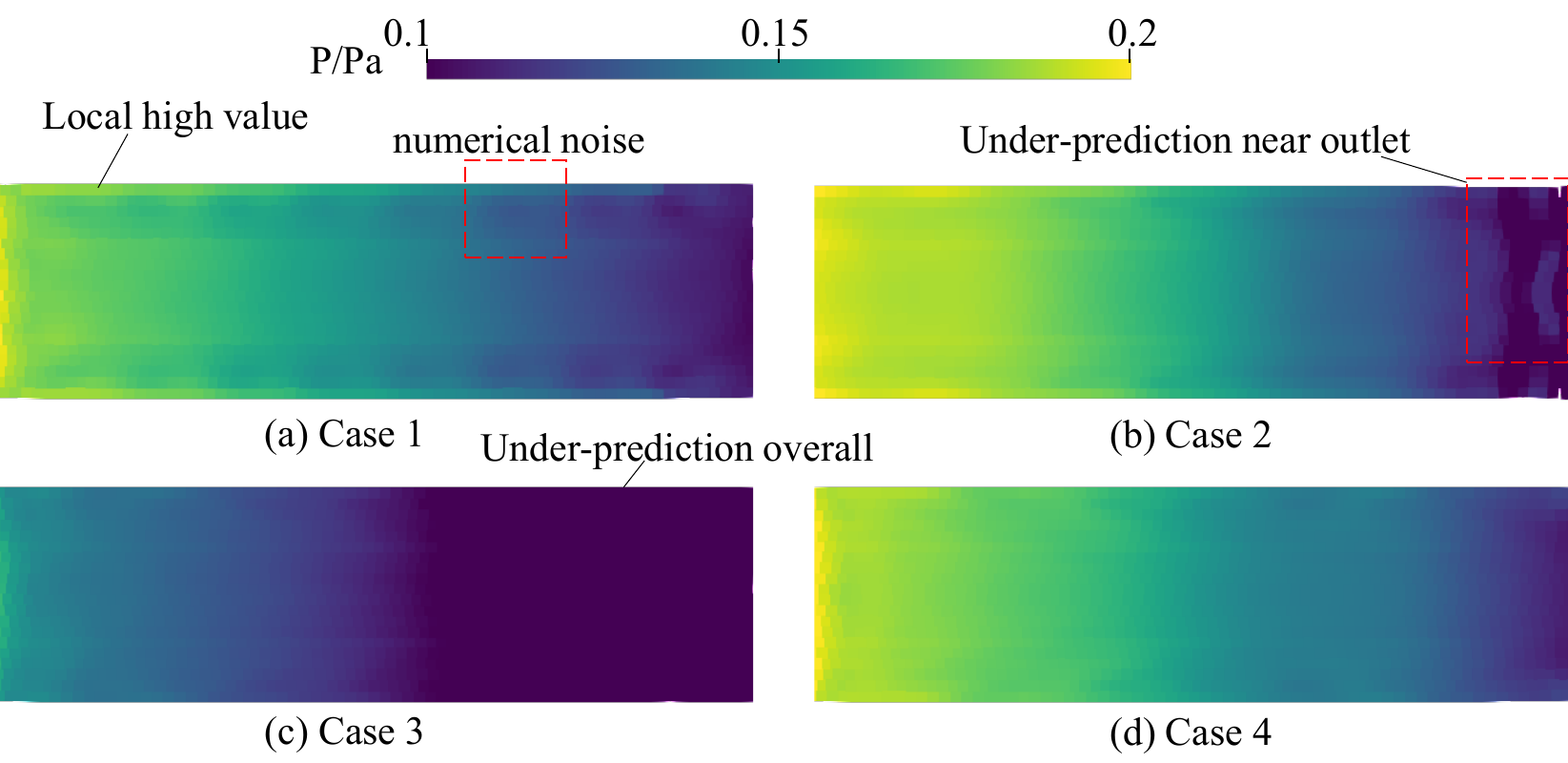}
	\caption
	{
		Laminar straight channel flow: the pressure fields simulated by the SPH method after achieving the steady state under the four cases.
	}
	\label{fig-lam-straight-contour-p-cases}
\end{figure}

The pressure contours for the four cases are presented in Fig. \ref{fig-lam-straight-contour-p-cases}, and the condition of each case is concluded in Table \ref{tab:cases-obc}.

In Case 1, although the pressure distribution is reasonably good, periodically numerical noise appears near the wall boundary, and the wall-nearest fluid particles suffer a local high value.
In Case 2, with the RKGC, the consistency is improved and the numerical noise near wall disappears, however, without an appropriate treatment for the correction matrix, the flow is wrongly accelerated near the outlet due to the truncated high magnitude of the correction matrix on the open boundary.
In Case 3, with the mirror boundary condition, the inappropriate acceleration is handled, while whole pressure field suffers an overall under-prediction due to the inconsistency between the approximation schemes of the internal and open boundary pressure gradient.
In Case 4, with both the mirror boundary condition and the improved open boundary condition (Eq. \eqref{eq-pressure-gradient-obc-B-final}), the problems in the previous 3 cases are well addressed, and the pressure becomes smooth and continuous.

\begin{figure}[htb!]
	\centering
	\includegraphics[trim = 14.39cm 0cm 0cm 2cm, clip,width=0.6\textwidth]{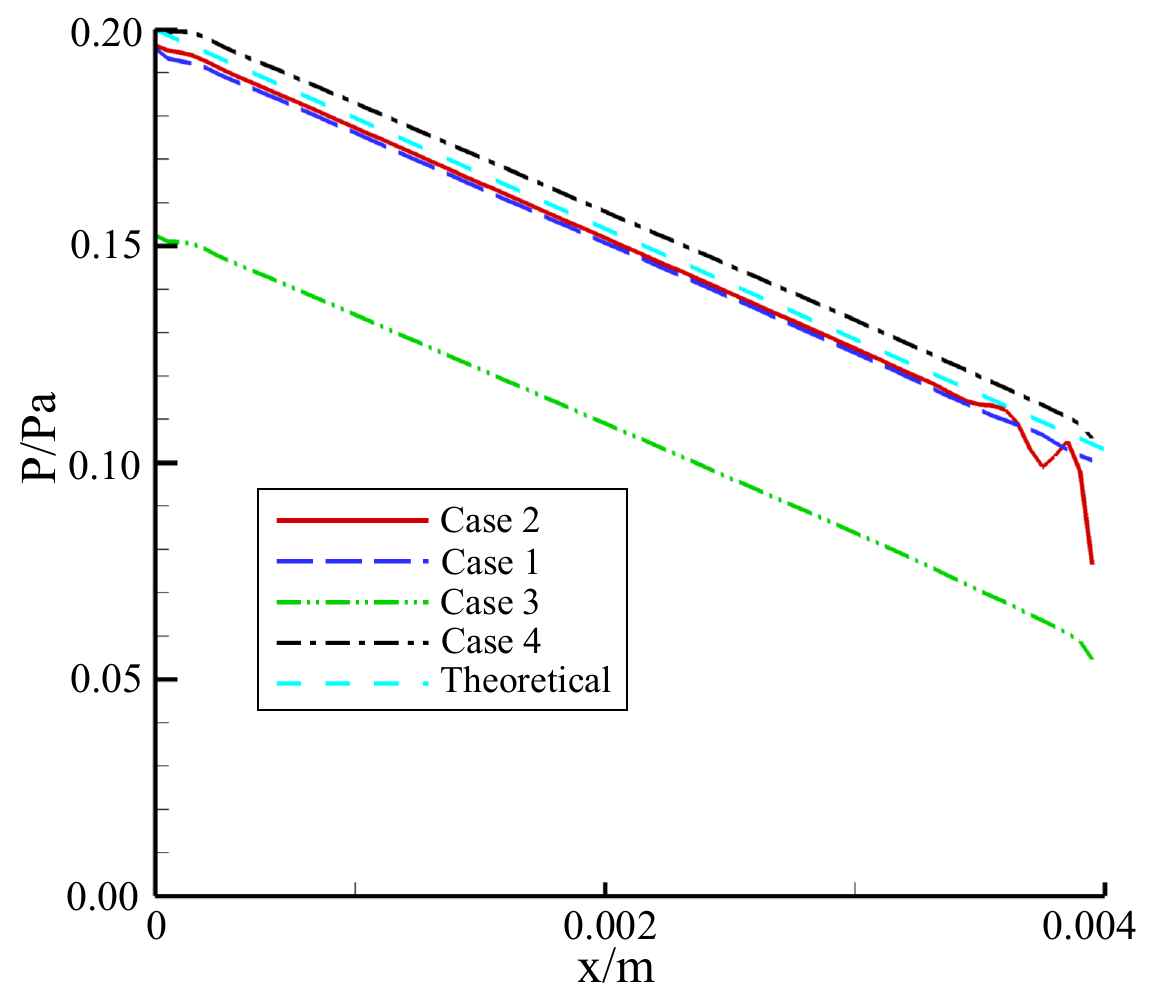}
	\caption
	{
		Laminar straight channel flow: the time-averaged centerline pressure profiles simulated by the SPH method after achieving the steady state under the four cases.
	}
	\label{fig-lam-straight-centerline-p-cases}
\end{figure}

\begin{figure}[htb!]
	\centering
	\includegraphics[trim = 14.39cm 0cm 0cm 2cm, clip,width=0.6\textwidth]{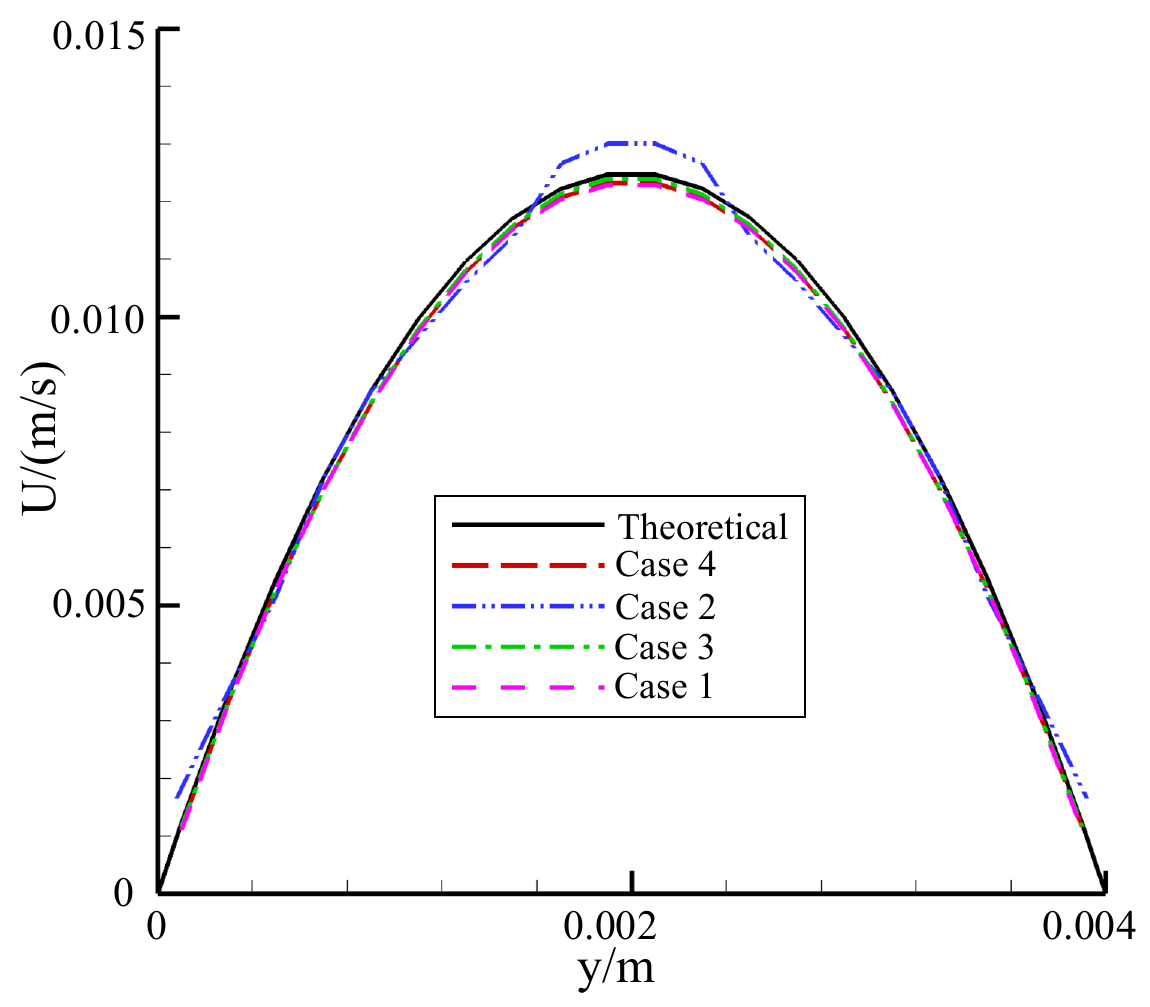}
	\caption
	{
		Laminar straight channel flow: the time-averaged cross-sectional velocity profiles at the outlet simulated by the SPH method after achieving the steady state under the four cases.
	}
	\label{fig-lam-straight-centerline-cross-vel-cases}
\end{figure}

The quantitative centerline data are shown in Fig. \ref{fig-lam-straight-centerline-p-cases}, and for this laminar case only, the theoretical inlet pressure is 0.2, as derived in Ref. \cite{zhang2025dynamical}.
The under-predictions in Case 2 and Case 3 are clearly presented.
Comparing Case 1 with Case 4, we find that the original treatment gently under-predicts the pressure while the improved one slightly over-predicts this value.
However, it is the pressure gradient which effects in the momentum equation, and for Case 1 and Case 4, the pressure gradients agree well with the theoretical value.
Therefore, the cross-sectional velocity profiles at the outlet achieve a satisfactory agreement, except Case 2, as shown in Fig. \ref{fig-lam-straight-centerline-cross-vel-cases}.

\begin{table}[h!]
	\centering
	\caption{Summary of the four test cases for the open boundary condition treatments.}
	\label{tab:cases-obc}
	\begin{tabularx}{\linewidth}{c>{\raggedright\arraybackslash}X}
		\toprule
		Case & Description                                                                                                   \\
		\midrule
		1    & Original VIPO: open boundary condition proposed in \cite{zhang2025dynamical}.                                 \\
		2    & VIPO $+$ unmodified RKGC: direct application of the correction technique in \cite{zhang2024towards}.          \\
		3    & VIPO $+$ corrected RKGC: RKGC with mirror boundary condition incorporated into the correction matrix.         \\
		4    & Improved VIPO: VIPO with both the mirror boundary condition and Eq.~\eqref{eq-pressure-gradient-obc-B-final}. \\
		\bottomrule
	\end{tabularx}
\end{table}


As for the turbulent simulation, the Reynolds number is 20000.
The solutions obtained from the finite difference method (FDM)\cite{wang2025weakly} are imposed on the inlet to accelerate the development of flow.
The inlet bulk velocity is 1 and the outlet pressure is 0.
The same four cases, as concluded in Table \ref{tab:cases-obc}, of the open boundary treatments are considered.

\begin{figure}[htb!]
	\centering
	\includegraphics[trim = 3.41cm 0cm 0cm 10.49cm, clip,width=1.0\textwidth]{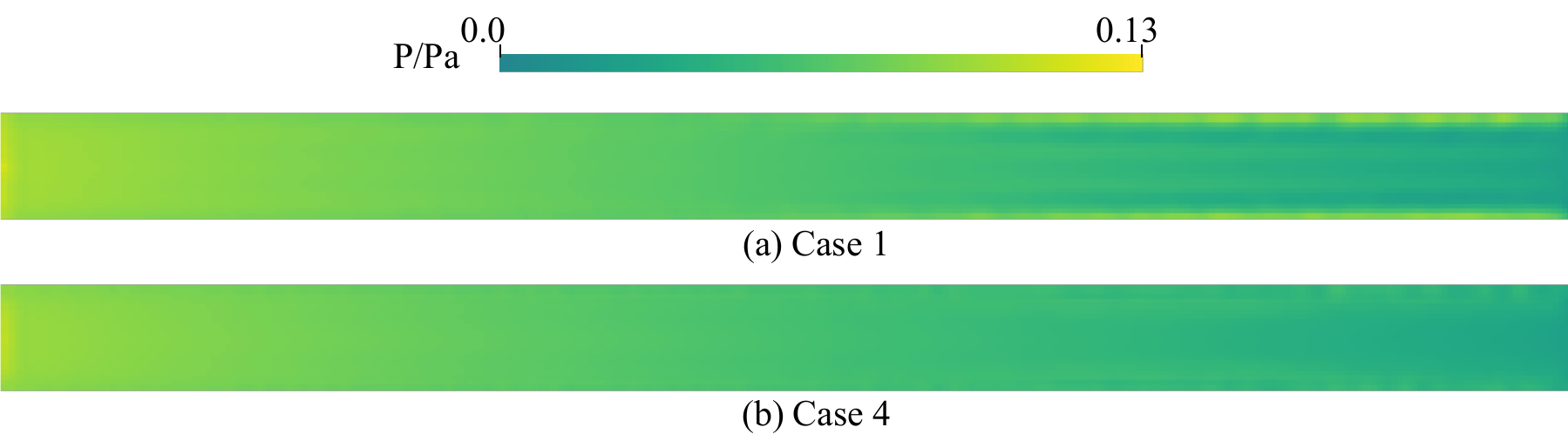}
	\caption
	{
		Turbulent straight channel flow: the time-averaged pressure contours simulated by the SPH method after achieving the steady state under the two cases.
	}
	\label{fig-turb-straight-contour-p-cases}
\end{figure}

The pressure contours are shown in Fig. \ref{fig-turb-straight-contour-p-cases}.
Since the characteristics of the pressure fields of Case 2 and Case 3 are very similar to those of the laminar simulations, we only present the contours of Case 1 and Case 4.
Without the improvement proposed in this work, an obvious high pressure region appears near the wall, becoming particularly severe when approaching the outlet.
In contrary, by using the improved open boundary condition, the pressure field becomes smooth and consistent.

\begin{figure}[htb!]
	\centering
	\includegraphics[trim = 2.99cm 0cm 0cm 5.67cm, clip,width=1.0\textwidth]{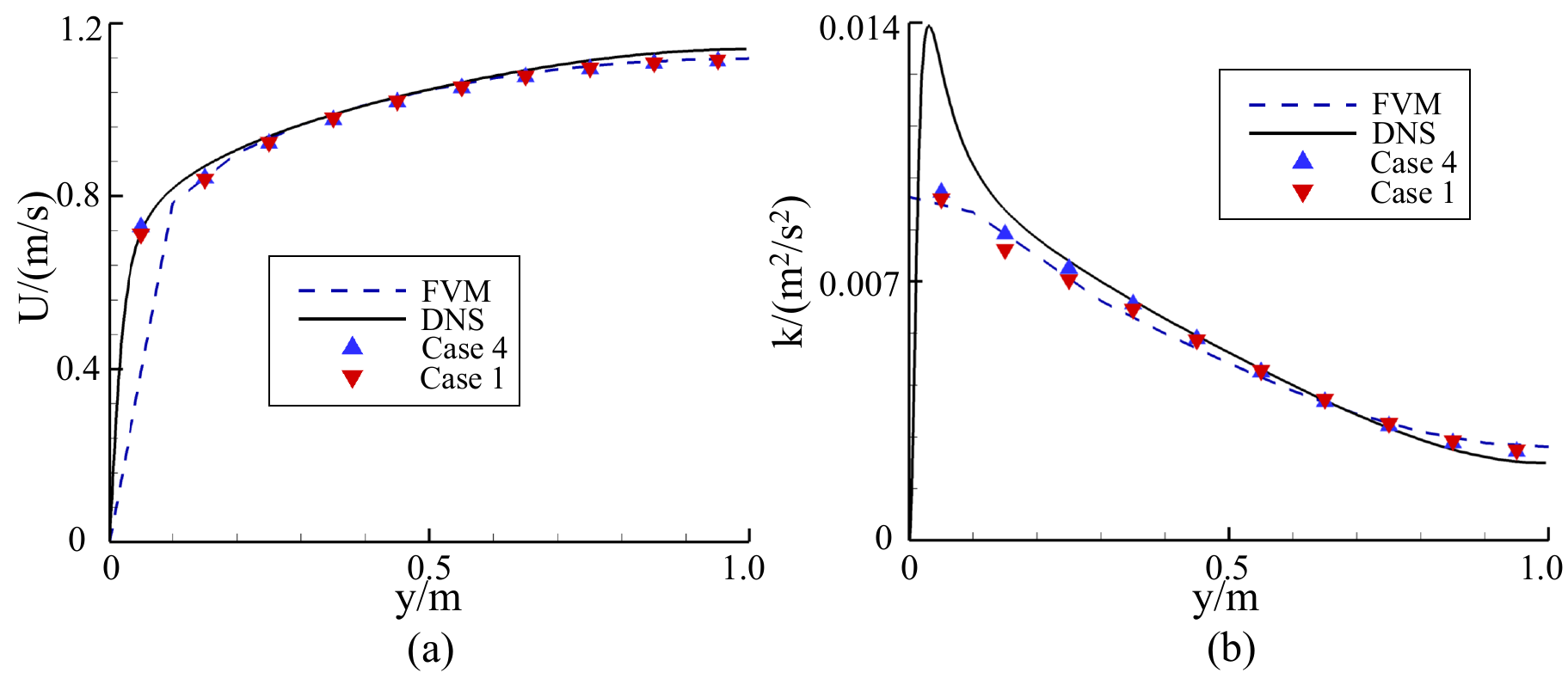}
	\caption
	{
		Turbulent straight channel flow: the time-averaged cross-sectional (a) velocity and (b) turbulent kinetic energy profiles at the outlet simulated by the SPH method after achieving the steady state under the two cases.
	}
	\label{fig-turb-straight-cross-k-cases}
\end{figure}

The cross-sectional velocity and turbulent kinetic energy profiles on the outlet are shown in Fig. \ref{fig-turb-straight-cross-k-cases}.
Although the results calculated under both the two cases agree well with those from the finite volume method\cite{wang2025weakly} and direct numerical simulation (DNS)\cite{lee2015direct}, an under-estimation is observed for Case 1 near the wall which may be due to the local high pressure observed in Fig. \ref{fig-turb-straight-contour-p-cases}.

%
\subsection{Flow through a U-shape channel}
\begin{figure}[htb!]
	\centering
	\includegraphics[trim = 8.04cm 0cm 0cm 6.65cm, clip,width=1.0\textwidth]{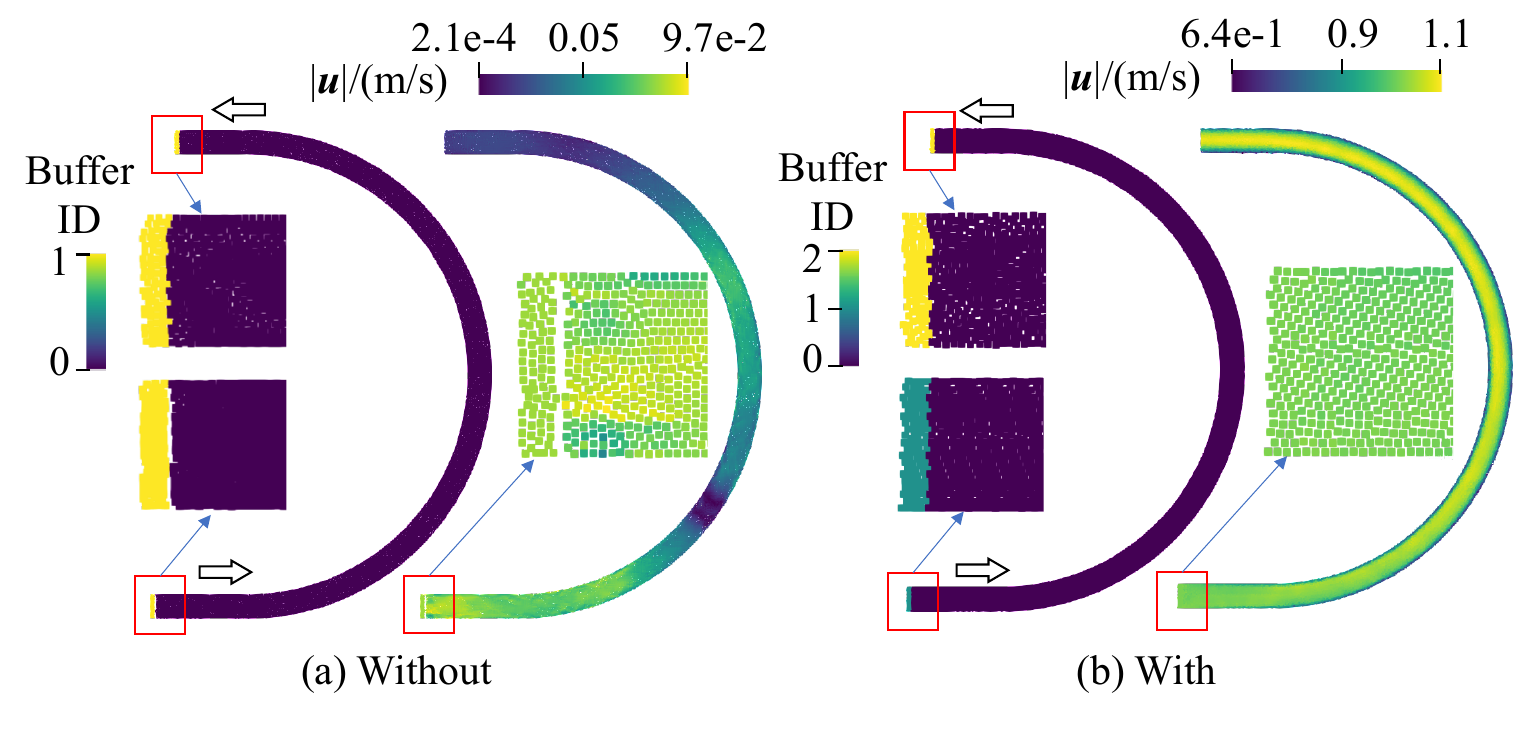}
	\caption
	{
		Turbulent U-shape channel flow: the buffer distribution and velocity contours simulated by the SPH method (a) without or (b) with the proposed improvement.
	}
	\label{fig-curved-contour-vel-with-without}
\end{figure}

To validate the improvement on the buffer independence mentioned in Section \ref{sec-independence}, the flow through a U-shape channel case is simulated.
Only the turbulent condition is considered, since the laminar result is similar to that of the laminar straight channel case.
The geometry is shown in Fig. \ref{fig-curved-contour-vel-with-without} (a), where one segment near the inlet buffer is within the deleting region of the outlet.
The Reynolds number is 148400, and the uniform velocity inlet and zero pressure outlet boundary conditions are used.

\begin{figure}[htb!]
	\centering
	\includegraphics[trim = 14.64cm 0cm 0cm 1.97cm, clip,width=0.8\textwidth]{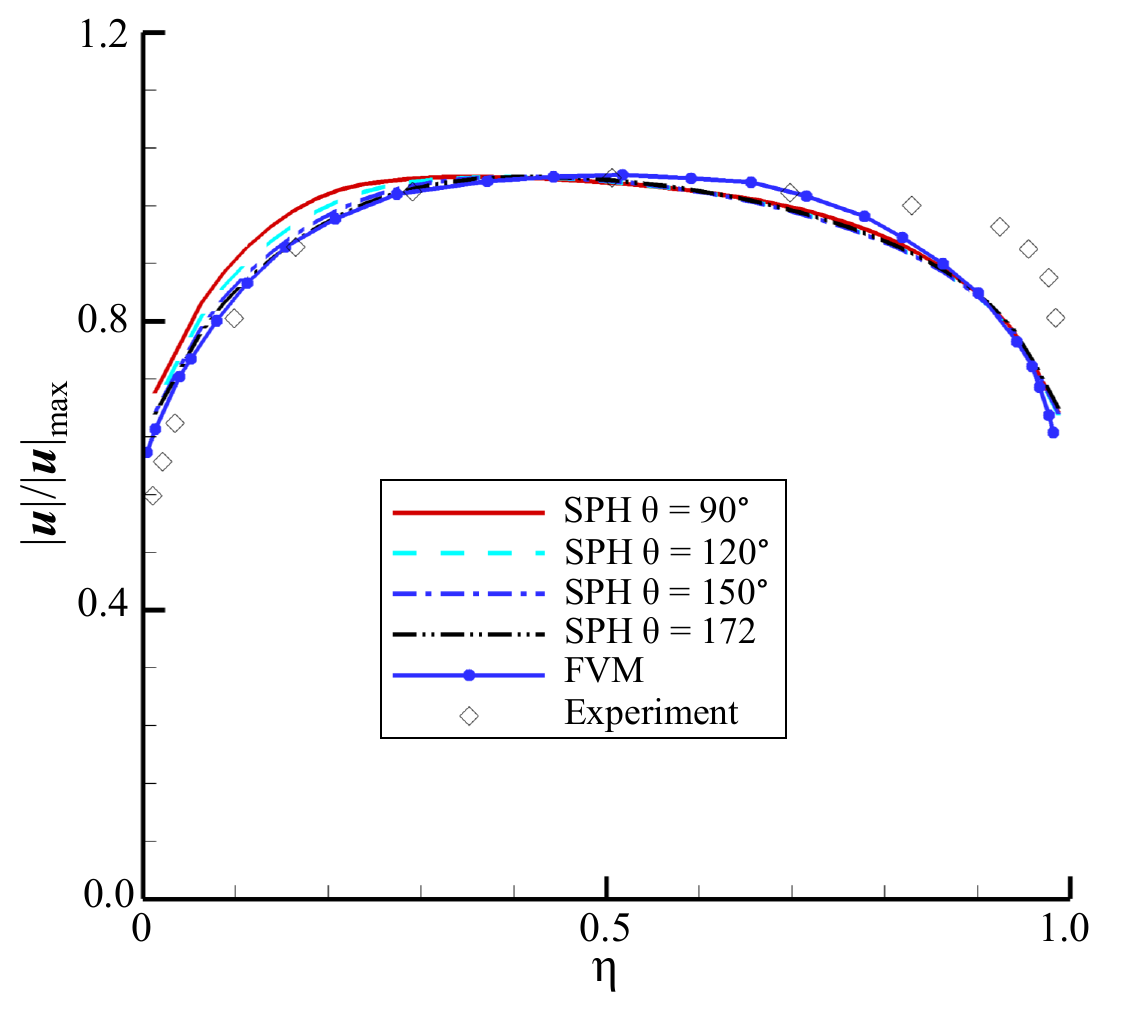}
	\caption
	{
		Turbulent U-shape channel flow: the comparison of the cross-sectional velocity profiles simulated by the SPH method at the four central angles, $\theta$, the finite volume method and from the experiment.
	}
	\label{fig-curved-cross-vel-k-combine}
\end{figure}

The velocity contours are shown in Fig. \ref{fig-curved-contour-vel-with-without}.
Without the proposed improvement, demonstrated in Fig. \ref{fig-curved-contour-vel-with-without} (a), the fluid particles near the inlet are unexpectedly deleted, and the simulation crashes.
In contrast, with the improvement, the wrong deletion is well avoided.
The quantitative data are shown in Fig. \ref{fig-curved-cross-vel-k-combine}.
For the SPH method, the velocity becomes fully-developed when the central angle is larger than 150$\circ$ and the profiles agree well with that obtained from the FVM\cite{pourahmadi1983prediction}.
Although compared with the experiment\cite{eskinazi1956investigation},the two numerical methods both under-predict the velocity values near the outer side of the curved channel due to the secondary flow.

%

\subsection{Plane jet}
As a typical benchmark case for examining the free-shear behavior in both laminar and turbulent \cite{white2006viscous} flow models, the plane jet represents an important flow configuration encountered in various engineering applications, such as paper manufacturing and rocket exhaust systems.
However, achieving accurate simulations of this case with particle-based methods remains challenging, particularly at high Reynolds numbers.

This may due to the three difficulties,
first, without the bidirectional buffer technique\cite{zhang2025dynamical} that can be regarded as the non-reflective far-field boundary condition, establishing a stable and symmetric potential flow region may be challenging.
Adopting the wall or fixed particles as the tank boundary may break the flow symmetry at high Reynolds number \cite{aristodemo2015sph}, and results in difficulty on obtaining the accurate centerline velocity.
Extending the transverse computational domain\cite{nazari2012numerical} is also a remedy but causes undesirable computational efforts, and most of the existing works focus on the submerged jet flow\cite{nazari2012numerical,de2020numerical}.
Second, the gentle and continuous backflow near the outlet further imposes challenges on the boundary condition of the particle-based method.
Since the vertical velocity does not vanish near the edge of the jet but instead is directed inwards or towards the jet, fluid is entrained across the boundary, causing the backflow.
The backflow continuously triggers the particle injection and causes the above-mentioned particle wandering problem.
Third, without the improvement on the buffer independence, the above-mentioned buffer interference problem could occur.

In this section, we first test the laminar plane jet case and compare the results with the analytical solutions\cite{white2006viscous}, then simulate the turbulent plane jet case without or with the RANS model.
It should be noted that even without incorporating a RANS model, the SPH method is capable of reproducing turbulence to some extent when the spatial resolution is sufficiently high\cite{adami2012simulating}, similar to the implicit large eddy simulation(LES)\cite{hu2011scale} method.
\subsubsection{Buffer setting}

\begin{figure}[htb!]
	\centering
	\includegraphics[trim = 21.03cm 0cm 0cm 5.37cm, clip,width=0.6\textwidth]{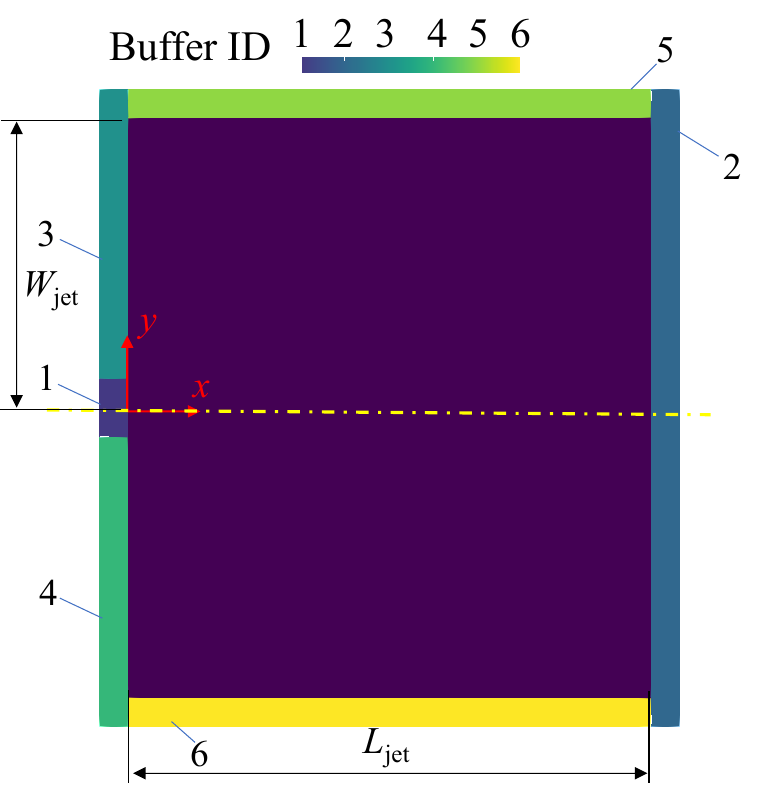}
	\caption
	{
		Plane jet: the size of the computational domain, buffer distribution and ID.
	}
	\label{fig-concept-jet-buffer}
\end{figure}

The geometry and the arrangement of the buffers are shown in Fig. \ref{fig-concept-jet-buffer}.
To exclude the influence of the wall, the six independent bidirectional buffers, indicated from 1 to 6, are used.
The boundary conditions imposed on the buffers are summarized in table \ref{tab-buffer-BC}.

\begin{table}
	\scriptsize
	\centering
	\caption{The boundary conditions on the buffers for the plan jet.}
	\renewcommand{\arraystretch}{1.2}
	\begin{tabular}{l c c c c}
		\hline
		Buffer No. & Type        & Velocity               & Pressure               & Turbulent quantities \\
		\hline
		1          & Inflow      & Uniform inlet velocity & Extrapolate from fluid & Freestream           \\
		2          & Bidirection & Extrapolate from fluid & Zero pressure          & Zero gradient        \\
		3, 4       & Inflow      & Zero velocity          & Extrapolate from fluid & Freestream           \\
		5, 6       & Bidirection & Extrapolate from fluid & Far-field pressure     & Zero gradient        \\
		\hline
	\end{tabular}
	\label{tab-buffer-BC}
\end{table}

\subsubsection{Laminar plane jet}
\begin{figure}[htb!]
	\centering
	\includegraphics[trim = 7.9cm 0cm 0cm 5.57cm, clip,width=1.0\textwidth]{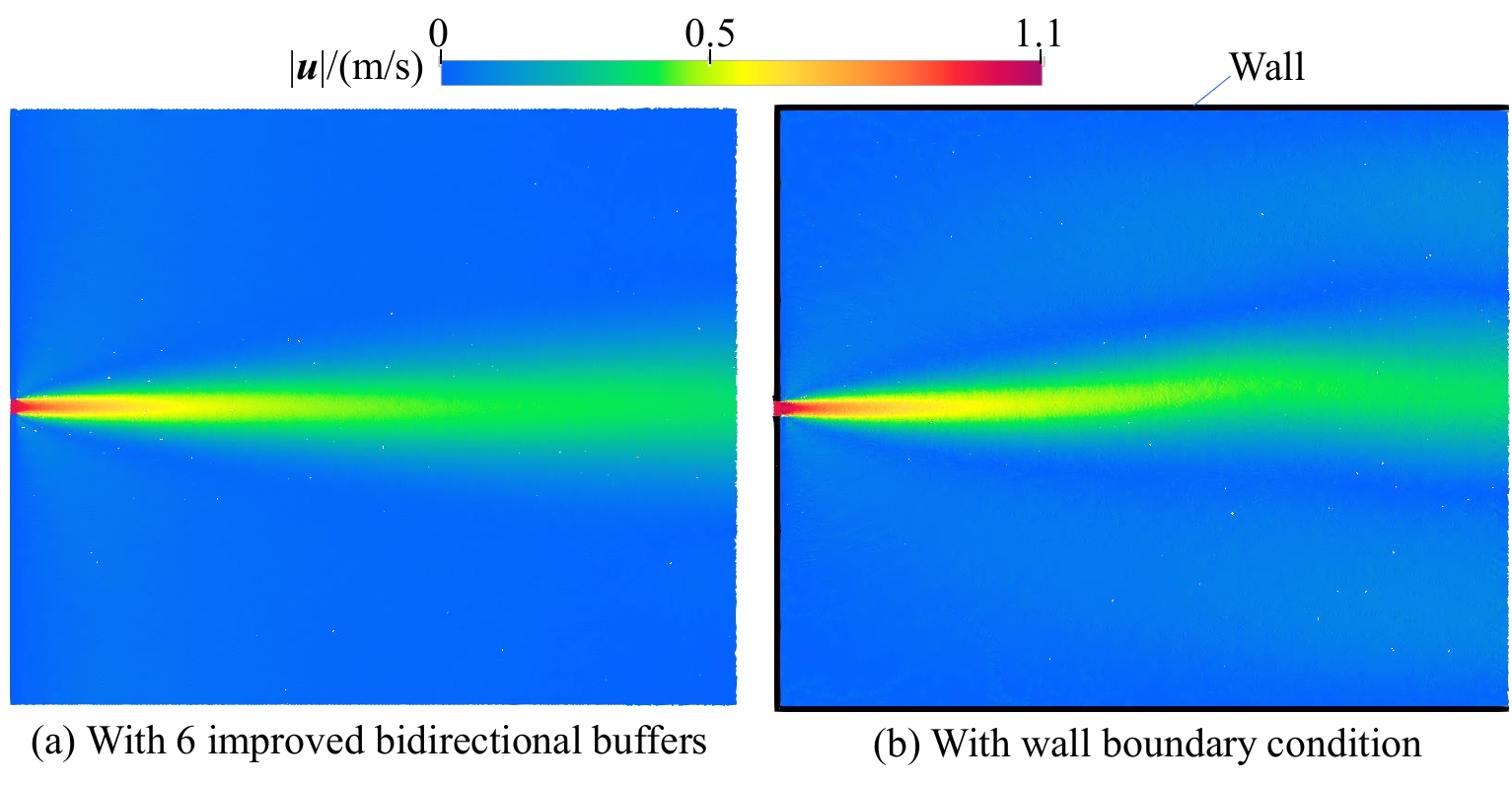}
	\caption
	{
		Laminar plane jet: the velocity contours at $Re=40$ and $N_f=10$, simulated by the SPH method (a) with  six improved bidirectional buffers and (b) with wall boundary condition.
	}
	\label{fig-lam-jet-Re40-100-with-without-wall}
\end{figure}

The Reynolds number, which is based on the inlet width and inflow velocity, is 40.
The resolution is defined as the number of fluid particles across the inlet width, denoted as $N_f$.
The inlet width $D_{jet}=2$, and the length and half width of the computational domain,$L_{jet}$ and $W_{jet}$, are $40D_{jet}$ and $20D_{jet}$, respectively.
Figure \ref{fig-lam-jet-Re40-100-with-without-wall} shows the velocity contours simulated by the SPH method with the improved six bidirectional buffers or with the wall boundary condition at the moderate Reynolds number, $Re=40$.

Using the wall boundary condition introduces disturbance on the velocity field, making the potential core region offset towards the upper wall.
The reason for the offset may be because that for the particle-based method with the wall dummy boundary, the particle near wall is keeping moving due the Lagrangian characteristic.
Therefore, the effective wall position is not strictly fixed, as the wall-adjacent fluid particles may slightly penetrate into or move away from the wall dummy interface, which can induce disturbances and break the flow symmetry.

Please note that this property will become more severe at a lower resolution, and hence using a higher resolution may mitigate this issue.
That may explain why the symmetric laminar plane jet in Aristodemo et al. \cite{aristodemo2015sph} was obtained under the wall boundary condition, as the lowest resolution used in their work($N_f=250$) is more than two hundred times higher than that in the present study.
However, such a high resolution is clearly not practical for engineering applications, and this issue may persist when the Reynolds numbers is high.

\begin{figure}[htb!]
	\centering
	\includegraphics[trim = 7.9cm 0cm 0cm 5.57cm, clip,width=1.0\textwidth]{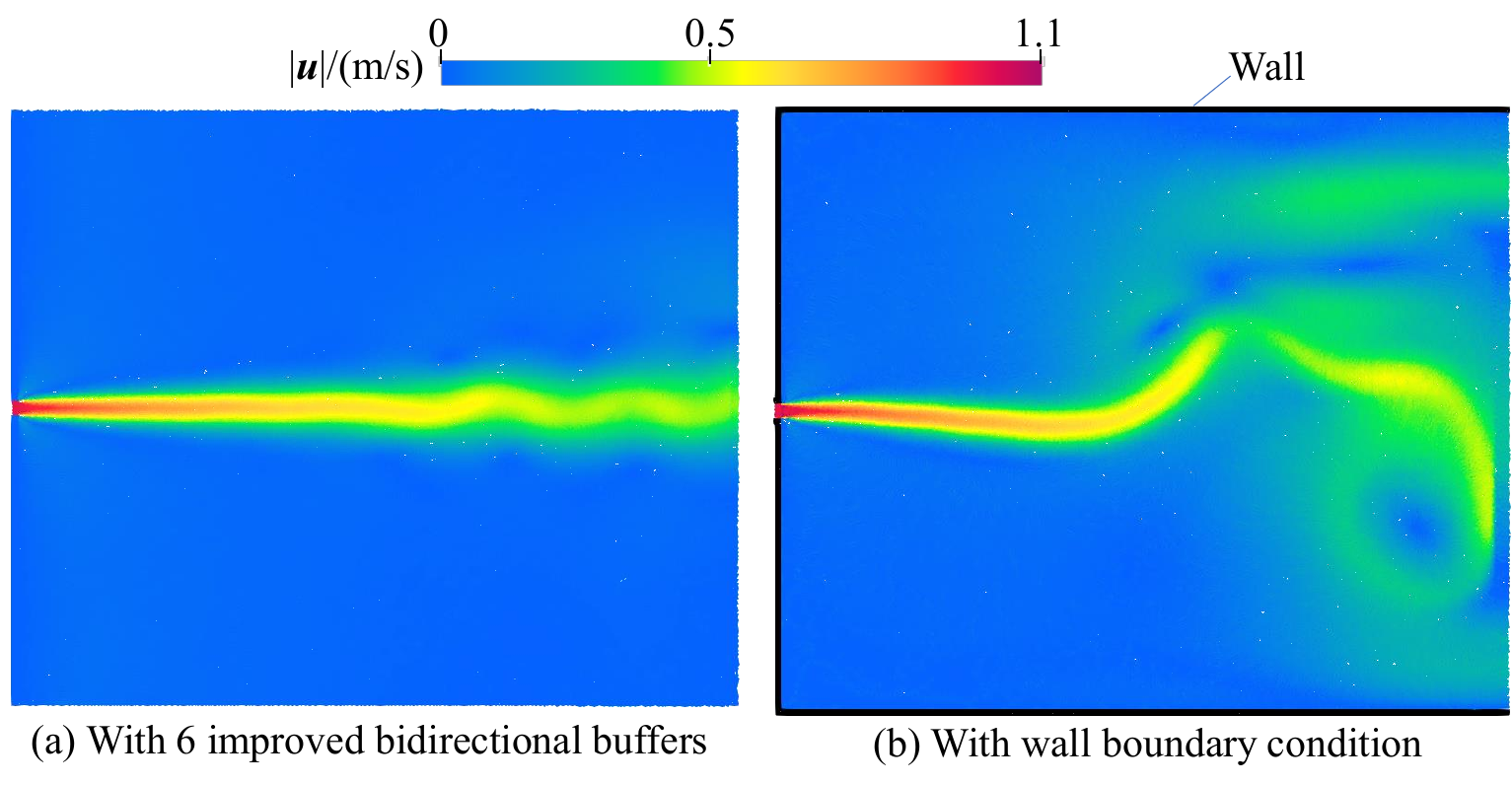}
	\caption
	{
		Laminar plane jet: the velocity contours at $Re=100$ and $N_f=10$, simulated by the SPH method (a) with  six improved bidirectional buffers and (b) with wall boundary condition.
	}
	\label{fig-lam-jet-Re100-real-with-without-wall}
\end{figure}

In contrary, using the improved bidirectional buffers avoids this problem at the moderate resolution, and a symmetric potential core region is obtained.
To further test the non-reflective property of the proposed open boundary condition, we increase the Reynolds number to 100, as shown in Fig. \ref{fig-lam-jet-Re100-real-with-without-wall}.
The instability appears near the outlet, but the stable potential core region still holds when the improved bidirectional buffers are used.
In contrary, the flow field becomes quite unstable due to the reflection from the wall, if the wall boundary condition is used.


\begin{figure}[htb!]
	\centering
	\includegraphics[trim = 15.48cm 0cm 0cm 4.14cm, clip,width=0.8\textwidth]{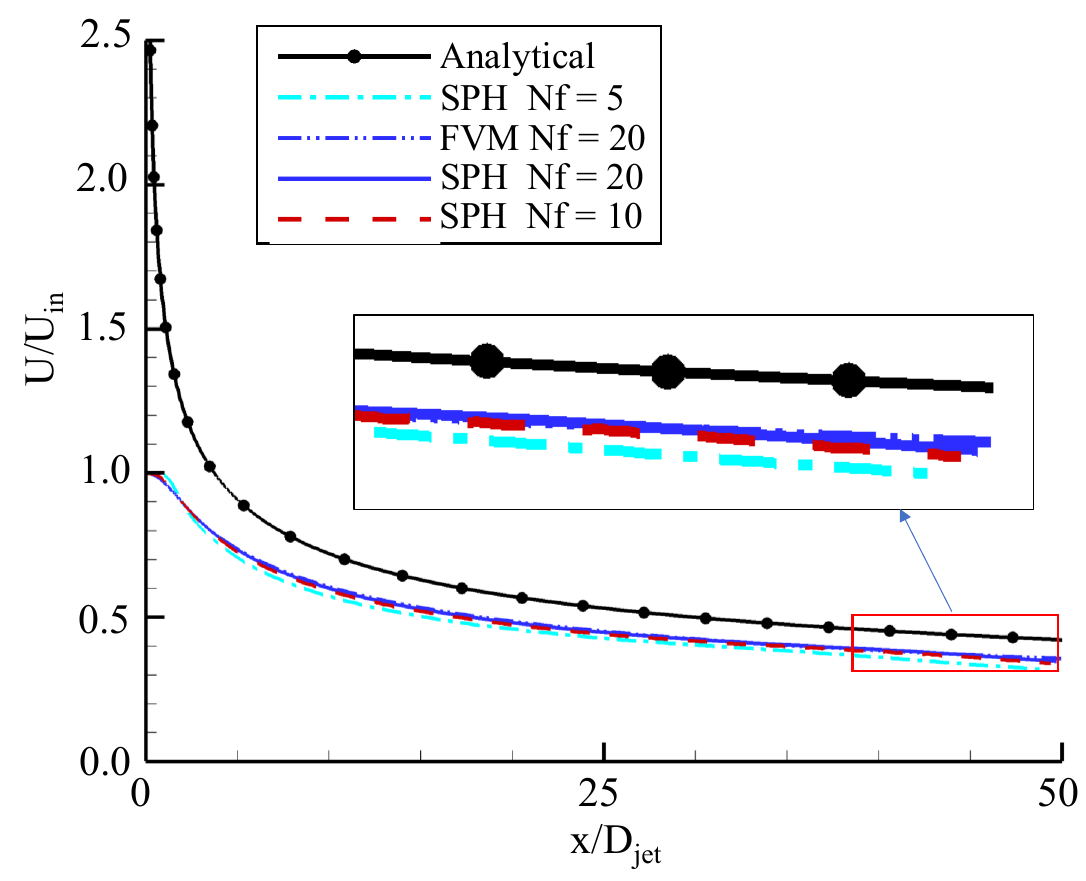}
	\caption
	{
		Laminar plane jet: the comparison of the centerline velocity profiles at $Re=40$ simulated by the FVM, SPH method, and from the analytical result.
	}
	\label{fig-lam-jet-Re40-centerline-vel}
\end{figure}

\begin{figure}[htb!]
	\centering
	\includegraphics[trim = 17.83cm 0cm 0cm 4.76cm, clip,width=0.8\textwidth]{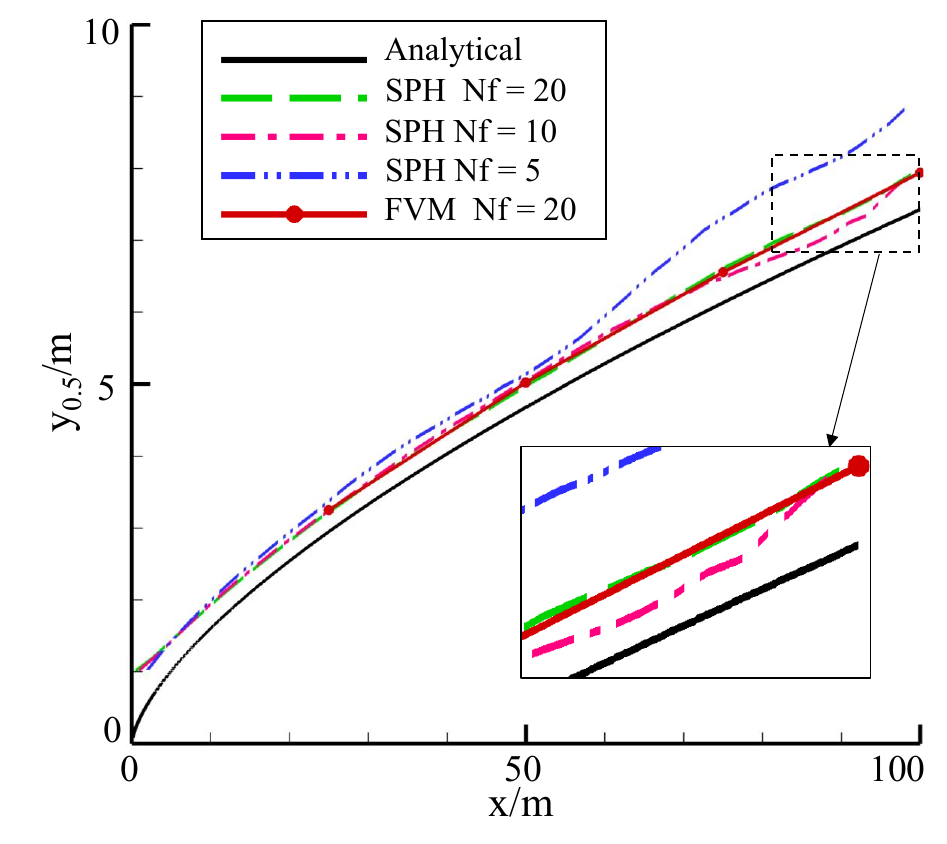}
	\caption
	{
		Laminar plane jet: the comparison of the spread rate at $Re=40$ simulated by the FVM, SPH method, and from the analytical result.
	}
	\label{fig-lam-jet-Re40-spread-rate}
\end{figure}

To conduct the quantitative comparison of the current SPH method, finite volume method (FVM) and the analytical solution, we test the centerline velocity\ref{fig-lam-jet-Re40-centerline-vel} and spreading width\ref{fig-lam-jet-Re40-spread-rate}.
All the SPH results are obtained by using the six improved bidirectional buffers, since it is difficult to compute meaningful data if the symmetry of the jet is not guaranteed when using the wall boundary condition.
The analytical solution is calculated according to the boundary layer theory\cite{schlichting2016boundary}.
As for the centerline velocity, as shown in Fig. \ref{fig-lam-jet-Re40-centerline-vel}, the convergence of the SPH method is satisfactory, and the results calculated by the SPH method agree well with that computed by the FVM at $N_f=20$.
Although both the two methods yield smaller centerline velocity near the inlet compared with the analytical solution.
This may be due to the drawback of the boundary theory, while the closing trend between the numerical and analytical results is obvious when approaching the outlet.

As for the spreading rate, at each cross-section across the mainstream direction, the distance from the centerline to the point whose velocity is half of the corresponding centerline velocity is defined as the $y_{0.5}$.
Figure \ref{fig-lam-jet-Re40-spread-rate} clearly presents the converging trend of the results calculated by the SPH method, and the two numerical methods agree well with each other.
Although both the FVM and SPH methods obtain higher spreading width compared with that of the analytical result, the slopes between the numerical and analytical solutions agree well with each other.

%

\subsubsection{Turbulent plane jet}
The Reynolds number, which is based on the inlet width and inflow velocity, is 20000.
To further validate the proposed improvement, the three different computational domains, namely small, compact and large domains, are tested.
The sizes of the domains are concluded in Table \ref{tab-turb-jet-size}, where the inlet width $D_{jet}$ is 2.

\begin{table}
	\centering
	\footnotesize
	\caption{The sizes of the computational domains}
	\renewcommand{\arraystretch}{1.2}
	\begin{tabular}{l c c}
		\hline
		Name of the domain & $L_{jet}/D_{jet}$ & $W_{jet}/D_{jet}$ \\
		\hline
		Small              & 10                & 5                 \\
		Compact            & 20                & 5                 \\
		Large              & 40                & 20                \\
		\hline
	\end{tabular}
	\label{tab-turb-jet-size}
\end{table}

\begin{figure}[htb!]
	\centering
	\includegraphics[trim = 9.64cm 0cm 0cm 2.97cm, clip,width=1.0\textwidth]{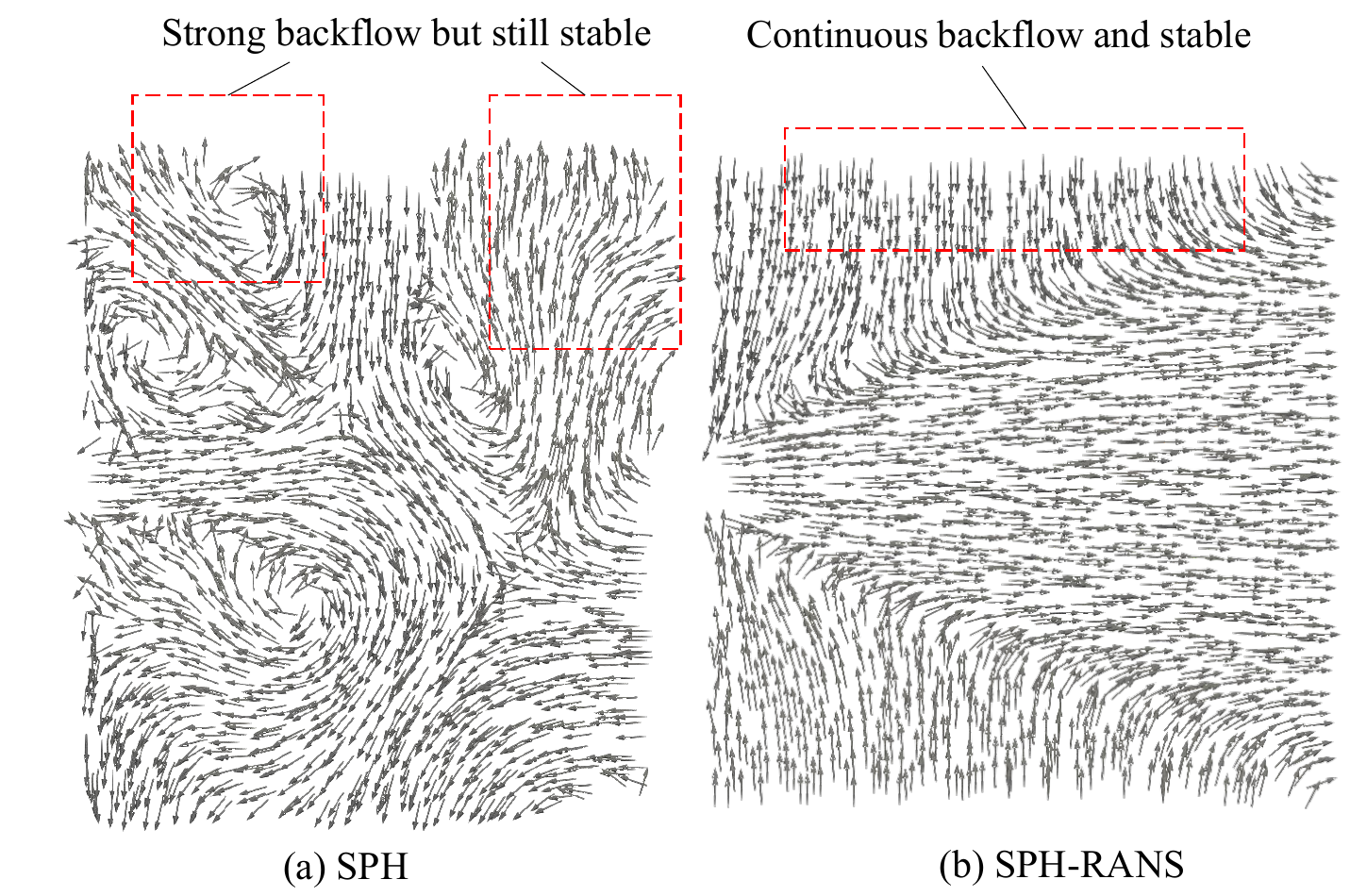}
	\caption
	{
		Turbulent plane jet: the velocity vector fields calculated by the SPH and SPH-RANS methods with the small computational domain, please note that the vector only represents the direction.
	}
	\label{fig-turb-jet-compact-Re20000-with-out-vortex}
\end{figure}

Firstly, to test the ability of the proposed improvement on handling the strong backflow, we simulate the small computational domain with and without the RANS model, respectively, as shown in Fig. \ref{fig-turb-jet-compact-Re20000-with-out-vortex}.
Please note that since the Riemann-based SPH method used in this work is intrinsically similar to the implicit large eddy simulation (ILES)\cite{hu2011scale}, directly using the Riemann based SPH method will lead to very unsteady velocity field due to the lack of sufficient resolution to resolve all the vortices, as shown in Fig. \ref{fig-turb-jet-compact-Re20000-with-out-vortex} (a).
However, although some strong backflow appears near the open boundary, the improved open boundary condition can still handle it, preventing the simulation from crashing.
While with the original bidirectional buffer, the simulation crashes immediately.

After adding the RANS model, as shown in Fig. \ref{fig-turb-jet-compact-Re20000-with-out-vortex} (b), the flow field becomes steady due to the eddy viscosity, and the continuous backflow appears near the upper and bottom buffers.
The simulation is still quite stable after introducing the improvements.

\begin{figure}[htb!]
	\centering
	\includegraphics[trim = 10.09cm 0cm 0cm 4.01cm, clip,width=1.0\textwidth]{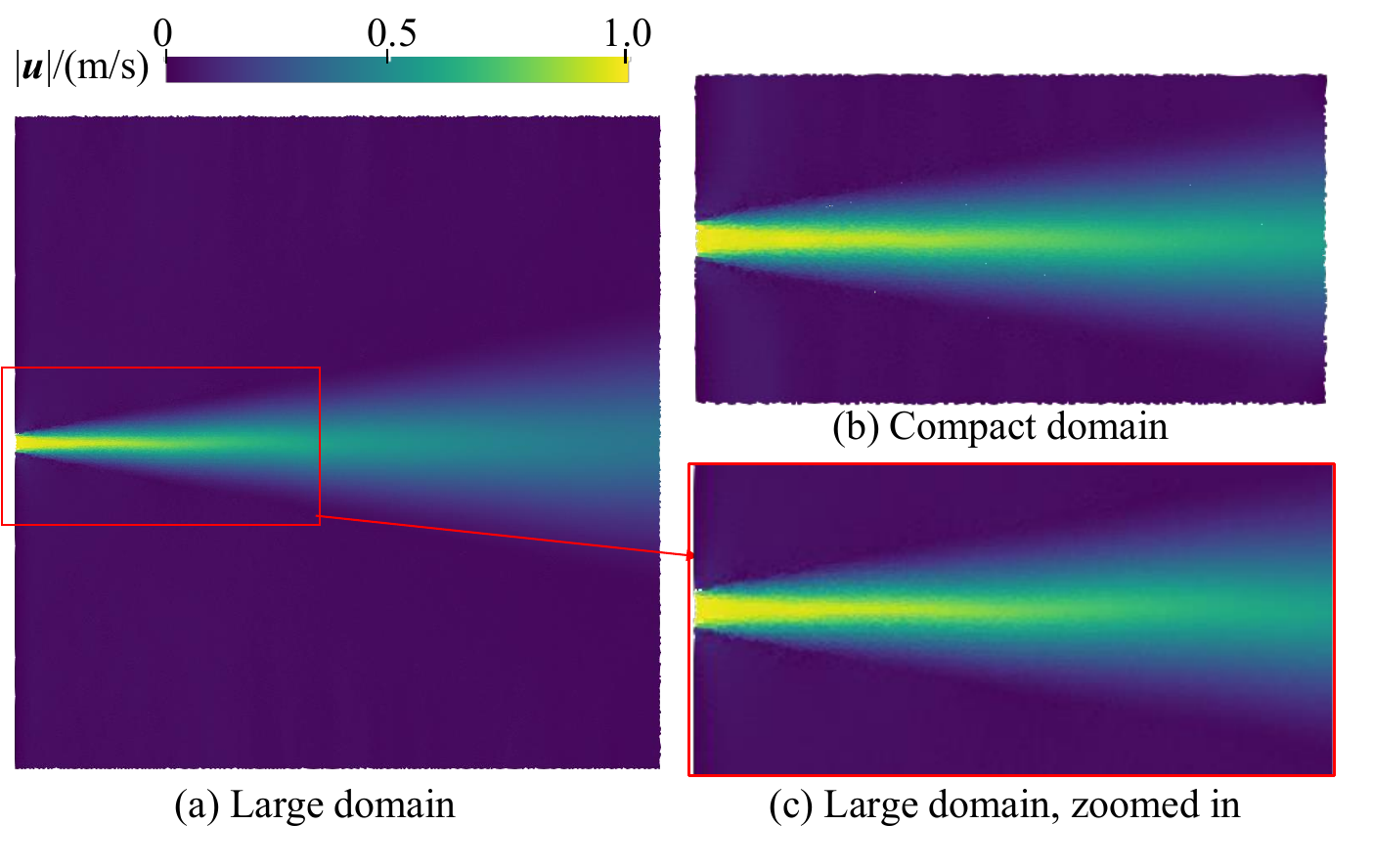}
	\caption
	{
		Turbulent plane jet: the velocity contours simulated by the SPH-RANS method at $N_f=10$ under the two different computational domains.
	}
	\label{fig-turb-jet-contour-domain}
\end{figure}

\begin{figure}[htb!]
	\centering
	\includegraphics[trim = 10.09cm 0cm 0cm 4.01cm, clip,width=1.0\textwidth]{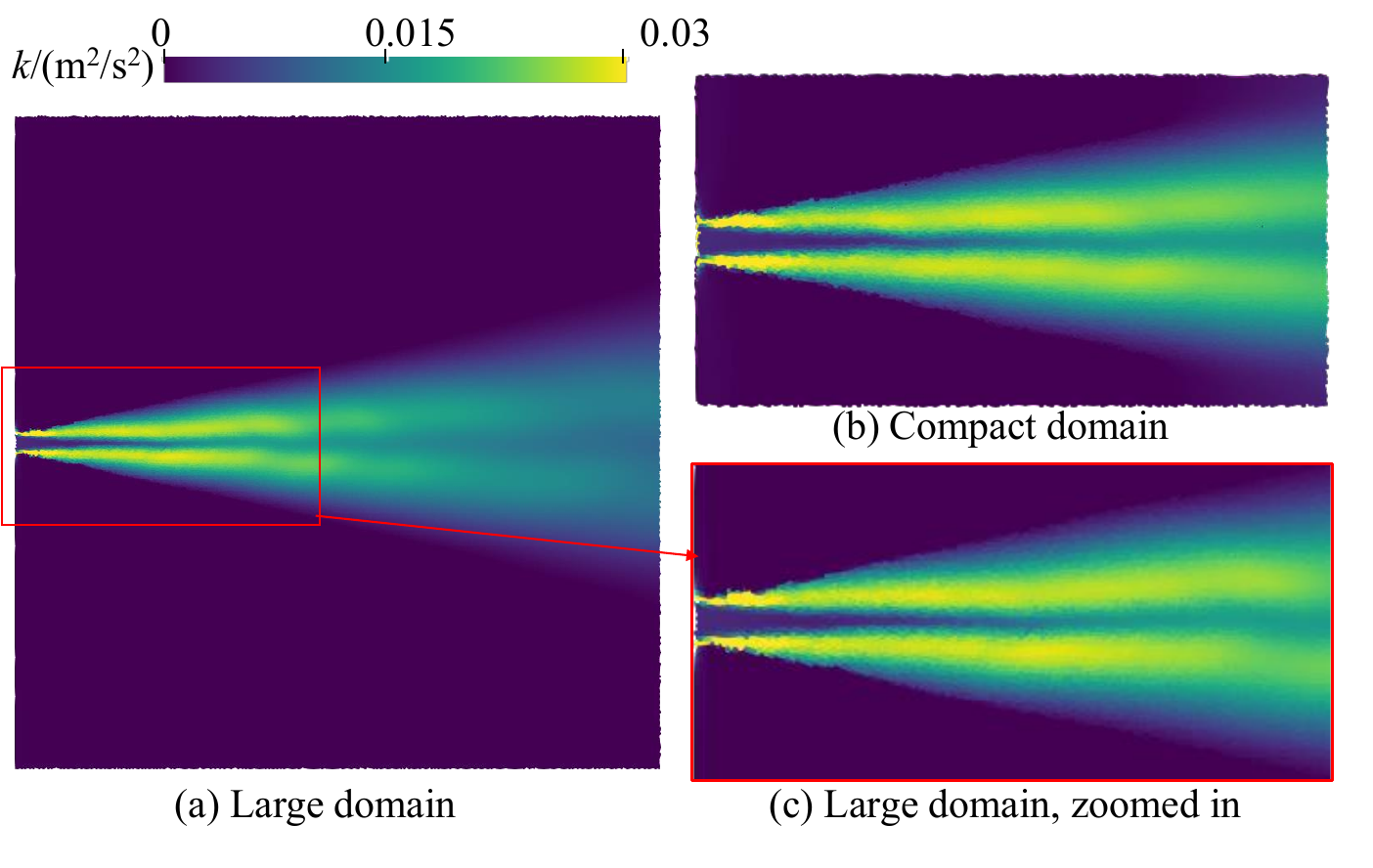}
	\caption
	{
		Turbulent plane jet: the turbulent kinetic energy contours simulated by the SPH-RANS method at $N_f=10$ under the two different computational domains.
	}
	\label{fig-turb-jet-k-contour-domain}
\end{figure}

The velocity contours calculated by the SPH-RANS method are shown in Fig. \ref{fig-turb-jet-contour-domain}.
The large (Fig. \ref{fig-turb-jet-contour-domain}(a) and (c)) and compact (Fig. \ref{fig-turb-jet-contour-domain}(b)) domains are separately simulated.
Although for the simulation using the compact domain, the upper, bottom and outlet open boundaries are quite close to the potential core region, no disturbance from the open boundaries is observed.
The velocity contours of the potential core region simulated under the two domains agree well with each other.
The same characteristic can be found on the contours of the turbulent kinetic energy, as shown in Fig. \ref{fig-turb-jet-k-contour-domain}.


\begin{figure}[htb!]
	\centering
	\includegraphics[trim = 7.25cm 0cm 0cm 2.3cm, clip,width=1.0\textwidth]{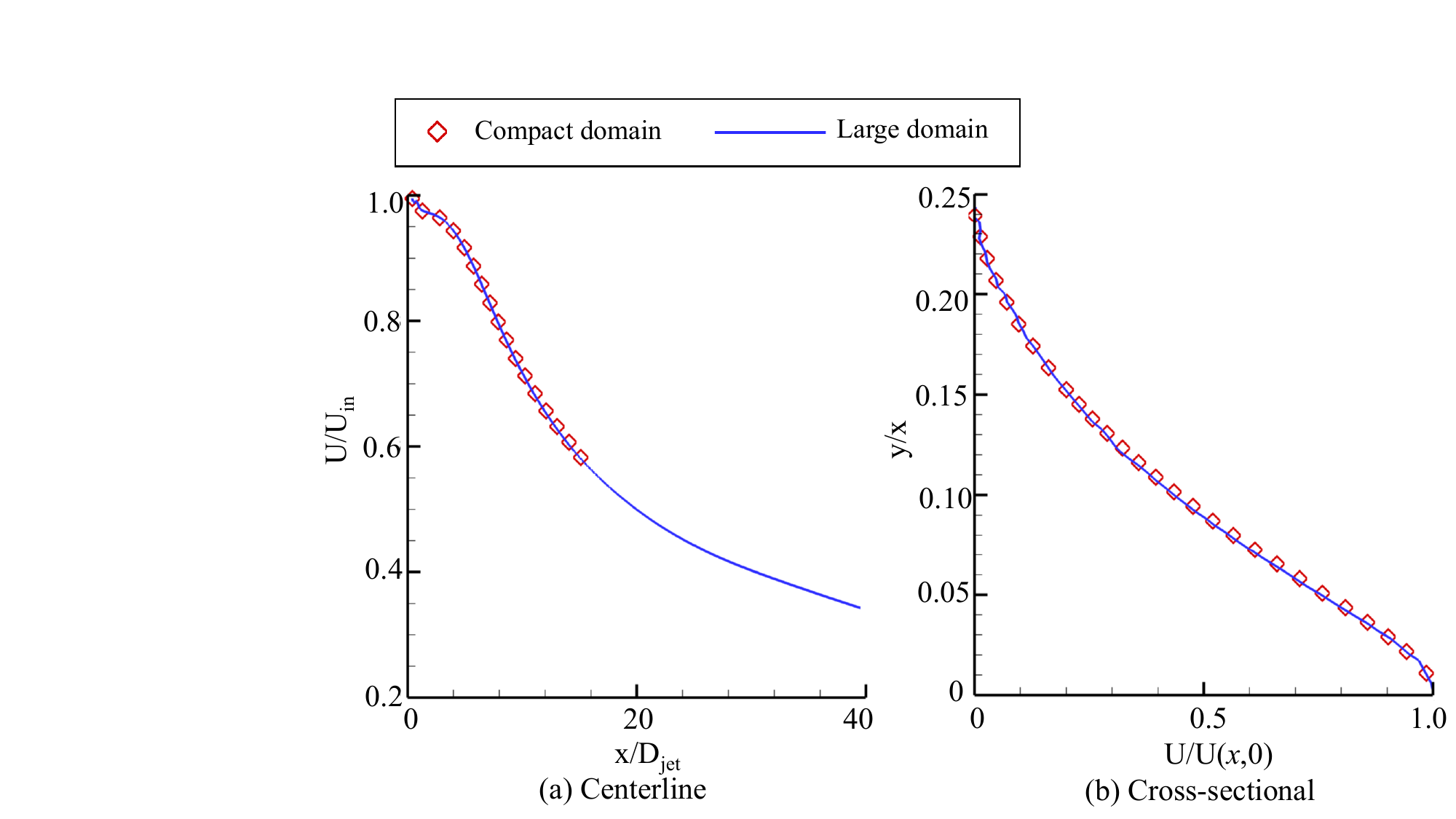}
	\caption
	{
		Turbulent plane jet: the comparison of the (a) centerline and (b) cross-sectional velocity profiles calculated by the SPH-RANS method under the two computational domains at $N_f=10$.
	}
	\label{fig-turb-jet-line-domain}
\end{figure}

\begin{figure}[htb!]
	\centering
	\includegraphics[trim = 12.97cm 0cm 0cm 1.78cm, clip,width=0.8\textwidth]{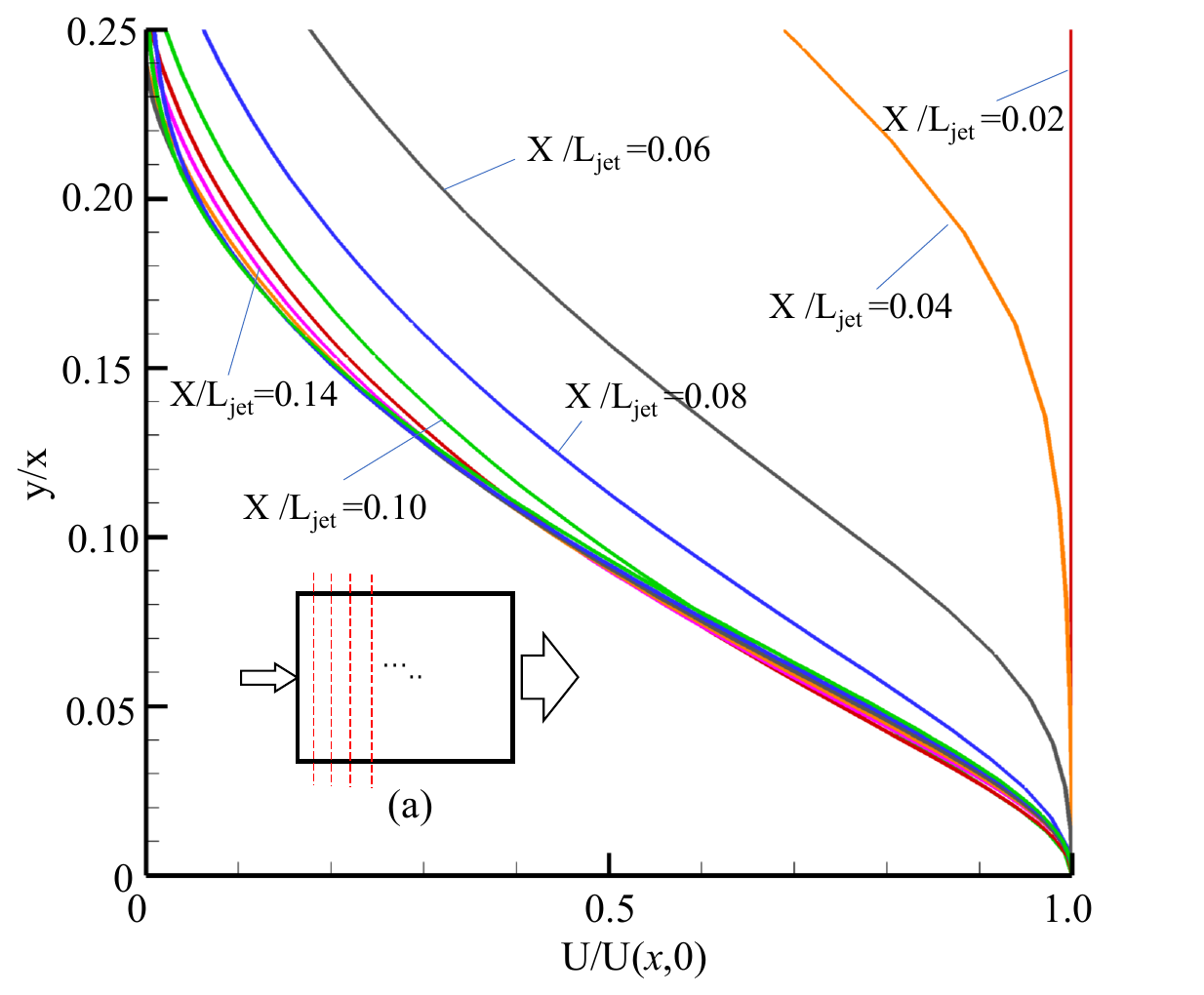}
	\caption
	{
		Turbulent plane jet: the cross-sectional velocity profiles at different $x$, the red dot lines shown in (a) stand for the monitoring positions.
	}
	\label{fig-turb-jet-selfsimilarity}
\end{figure}

To conduct the quantitative comparison, we compute the centerline and the cross-sectional velocity profiles, as illustrated in Fig. \ref{fig-turb-jet-line-domain}.
The cross-section is taken at $x=0.5L_{jet}$, while because of the self-similarity of this case (shown in Fig. \ref{fig-turb-jet-selfsimilarity}), any cross-sections which satisfy $x>0.14L_{jet}$ can yield almost the same cross-sectional velocity profiles.
The quantitative results indicate that the use of a compact domain introduces no significant difference, and the simulations performed in the two domains are in good agreement.


\begin{figure}[htb!]
	\centering
	\includegraphics[trim = 18.04cm 0cm 0cm 5.12cm, clip,width=0.8\textwidth]{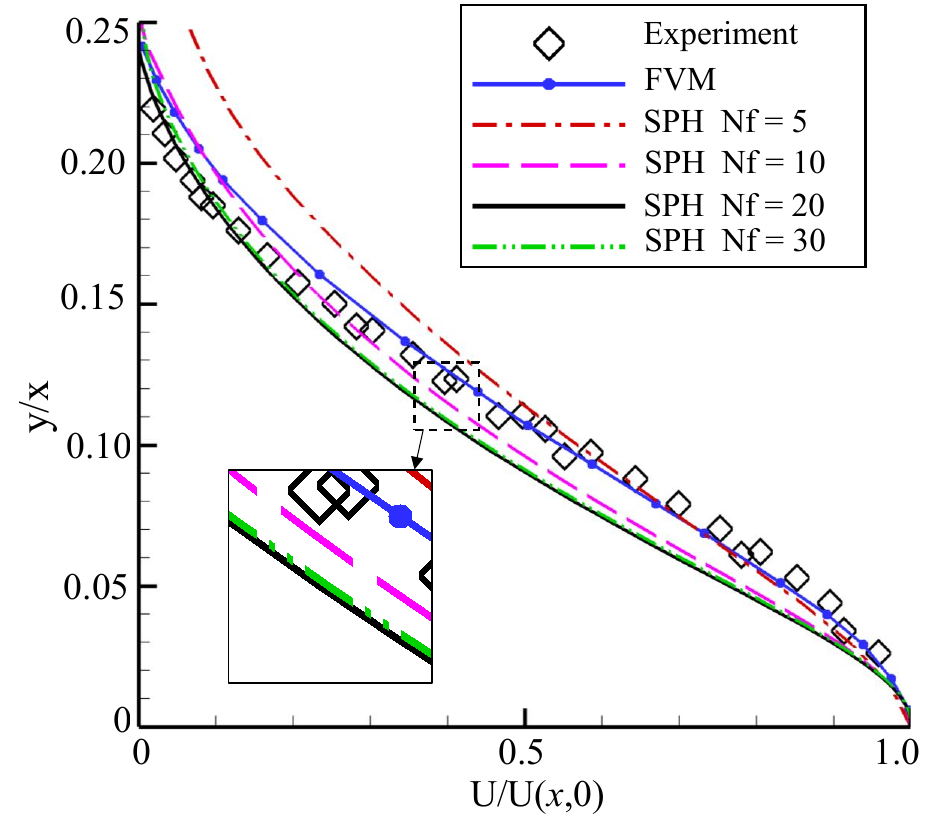}
	\caption
	{
		Turbulent plane jet: the comparison of the spread rates simulated by the FVM\cite{wilcox1998turbulence}, SPH method, and from the experiment\cite{bradbury1965structure}.
	}
	\label{fig-turb-jet-cross-line}
\end{figure}

Finally, we compare the results under the different resolutions with those calculated by the FVM\cite{wilcox1998turbulence} and from the experiment\cite{bradbury1965structure}, as shown in Fig. \ref{fig-turb-jet-cross-line}.
The results from the FVM are obtained with the same $k$-$\epsilon$ RANS model that is used in this work.
The SPH results show satisfactory convergence, and the cross-sectional velocity profile converges at $N_f=20$.
Compared with the FVM result, although the converged SPH result is smaller than that of the experiment, it shows a good agreement near the inlet.
Besides, it should be noted that, for the FVM result, the resolution and convergence test are not given in the reference\cite{wilcox1998turbulence}.

%

\subsection{Three-dimensional self-rotational micro-mixer}
\begin{figure}[htb!]
	\centering
	\includegraphics[trim = 13.29cm 0cm 0cm 7.29cm, clip,width=1.0\textwidth]{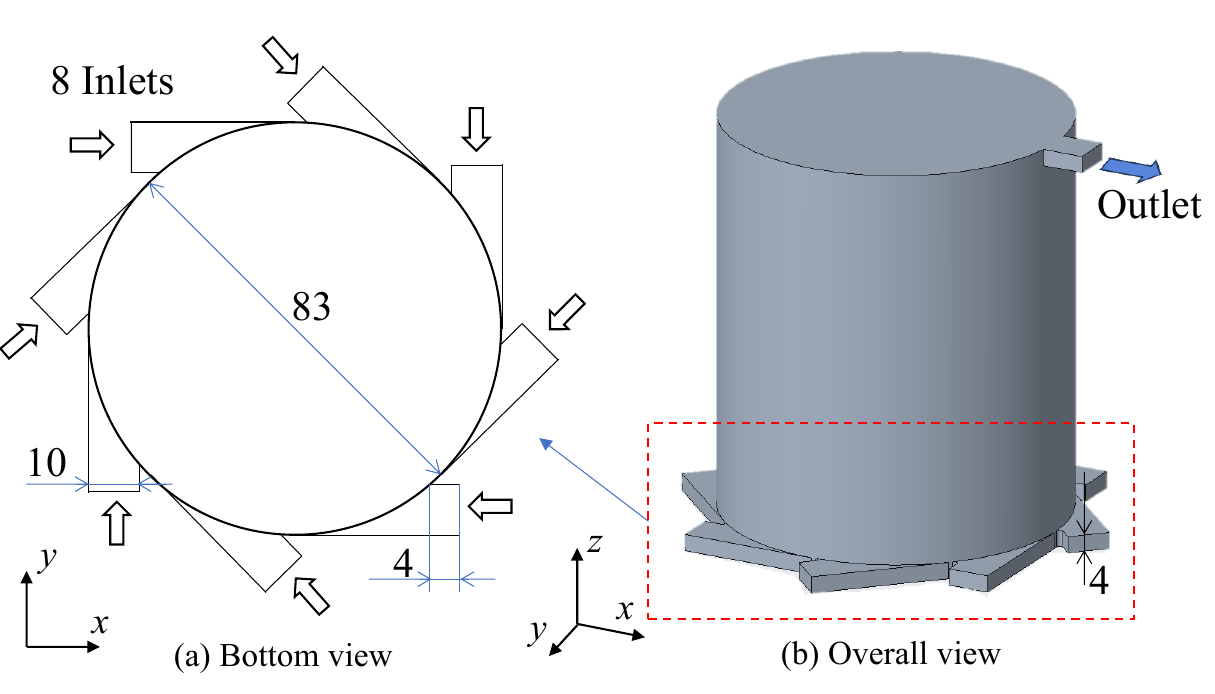}
	\caption
	{
		Self-rotational micro-mixer: the geometry and size.
	}
	\label{fig-mixer-geo}
\end{figure}

To test the proposed improvement on addressing the complex multiple in-outlets system, the three-dimensional self-rotational micro-mixer\cite{lin2005rapid} is simulated.
The geometry and size are shown in Fig. \ref{fig-mixer-geo}, the cross-sectional size of the outlet channel is the same as that of the inlet channel, and the total height is 100.
The self-rotation is triggered by the 8 inlets which inject fluid from the tangential direction of the cylinder chamber.

The Reynolds number, based on the hydraulic diameter of the inlet channel; therefore, only the laminar simulation is considered.
The resolution,$N_f$, is defined as the number of fluid particles across the height of the inlet channel.

\begin{figure}[htb!]
	\centering
	\includegraphics[trim = 6.22cm 0cm 0cm 2.36cm, clip,width=1.0\textwidth]{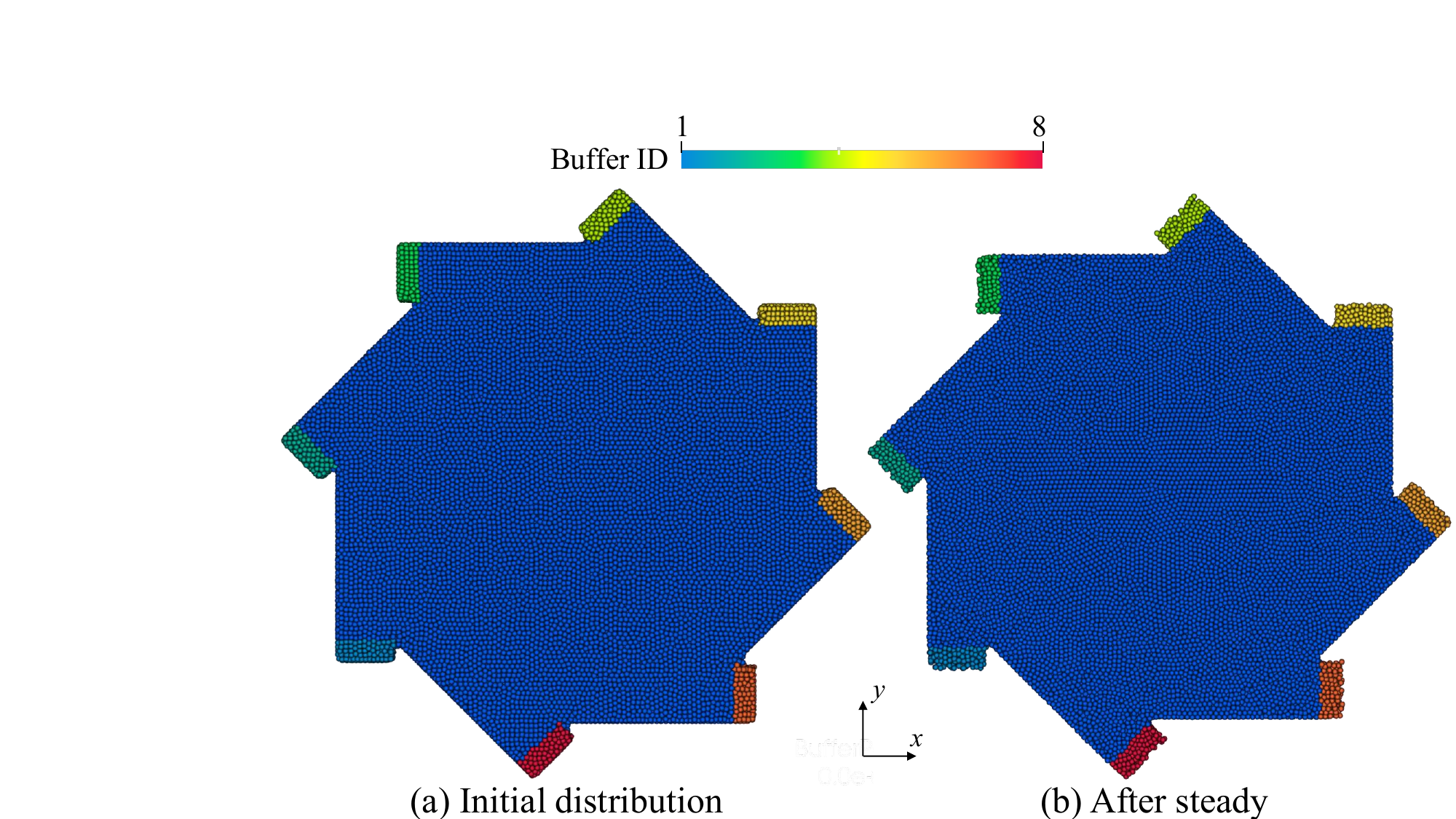}
	\caption
	{
		Self-rotational micro-mixer: the $Z=4$ cross-sectional particle distribution with the buffer ID (a) at initial state and (b) after steady.
	}
	\label{fig-mixer-buffer-id-steady}
\end{figure}

Figure \ref{fig-mixer-buffer-id-steady} (a) shows the initial particle distribution with the ID of each buffer at the cross-section where $z=4$.
Please note that the position of the buffer is deliberately placed very close to the fluid domain, so that the extreme condition is considered to fully validate the improvements.
The particle distribution in the buffers after the simulation achieves the steady state is shown in Fig. \ref{fig-mixer-buffer-id-steady} (b).
Each buffer is correctly tagged and the particles in the buffer are uniformly distributed, the particle interference and erroneous deletion are well avoided.

\begin{figure}[htb!]
	\centering
	\includegraphics[trim = 11.54cm 0cm 0cm 0.18cm, clip,width=1.0\textwidth]{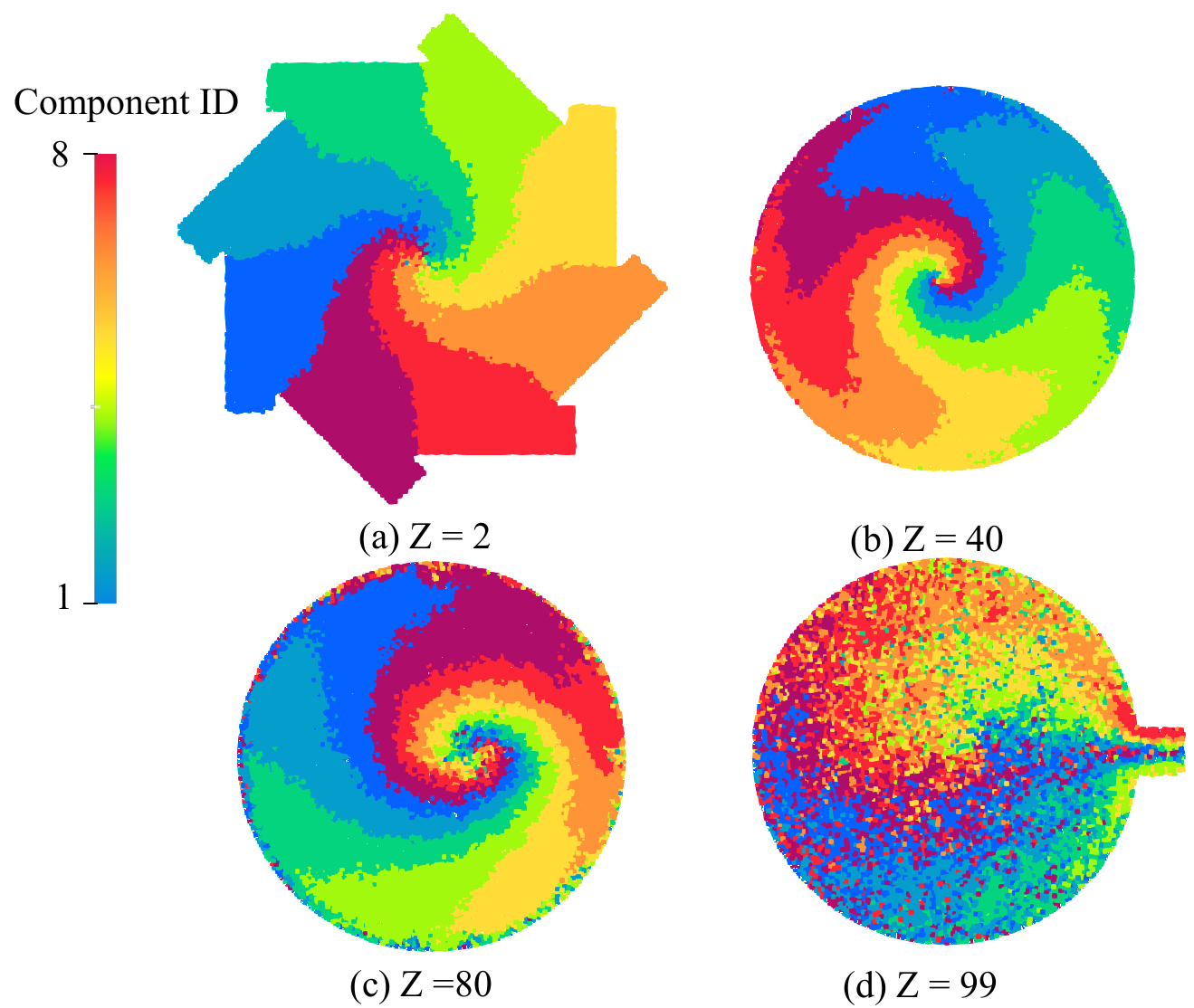}
	\caption
	{
		Self-rotational micro-mixer: the mixing performance at different cross-sections.
	}
	\label{fig-mixer-mixing-effect}
\end{figure}

Figure \ref{fig-mixer-mixing-effect} presents the mixing effect at the four cross-sections.
To clearly exhibit the mixing performance and prove that the improved bidirectional buffer works well under this extreme condition, the 8 kinds of immiscible fluids are injected from the 8 inlet channels.
At $Re = 17.1$, a distinct self-rotation is observed at $z = 2$, while the mixing remains relatively weak.
At $z = 40$, a clear rotational center emerges, and the eight fluid streams begin to mix with each other.
At $z = 80$, since there is only one outlet located near the top wall, the rotational center becomes offset, and the mixing intensity increases, particularly around the rotational core and the chamber wall.
At $z = 99$, the rotational center disappears, and the eight fluids are well mixed due to the outlet effect.

Please note that clearly capturing the interfaces between different fluids is an inherent advantage of the SPH method, which may be challenging for mesh-based methods to achieve.
In addition, only qualitative results are presented in this study, primarily to demonstrate the stability of the improved bidirectional buffer in the 3D simulation.
The quantitative comparison and convergence test will be conducted in the future.

%


\section{Conclusion}
\label{section-conclusion}
In this work, based on the latest WCSPH-RANS method, the open boundary treatment implemented in SPHinXsys is systematically introduced, and three improvements are proposed for complex open-boundary flows involving strong backflow.
The three improvements, focusing on the consistency, independence, and accuracy of the buffer-based open boundary condition, are comprehensively validated through a series of benchmark cases.

Firstly, the laminar and turbulent straight channel flows demonstrate that the accuracy improvement can effectively suppress the numerical pressure noise near the boundary, although a slight increase in background pressure is observed.
Secondly, the turbulent U-shaped channel flow validates the independence improvement, and the results show excellent agreement with those obtained from the finite volume method (FVM).
Thirdly, the laminar and turbulent plane jet cases thoroughly demonstrate the effectiveness of the proposed approach in handling strong backflow.
Satisfactory results are obtained within a compact computational domain, which is much smaller than that required by the FVM, indicating a substantial reduction in computational cost.

In addition, the three-dimensional self-rotational micro-mixer with a complex inlet/outlet configuration and extremely compact buffer arrangement is qualitatively tested. Each buffer region remains stable, and the expected self-mixing performance is successfully achieved.
Overall, the proposed improvements significantly enhance the robustness and applicability of the WCSPH-RANS framework for simulating complex open-boundary flows.

	%
	%
	{ \section*{Appendix A. The values of the coefficients}}
\label{appendix}
The coefficients of the $k$-$\epsilon$ model are listed in Table \ref{tab-coeff-ke}.

\begin{table}
	\scriptsize
	\centering
	\caption{Coefficients for the standard $k-\epsilon$ RANS model.}
	\begin{tabularx}{8.5cm}{@{\extracolsep{\fill}}lc}
		\hline
		Name of the coefficients & Value  \\
		\hline
		$C_1$                    & $1.44$ \\
		\hline
		$C_2$                    & $1.92$ \\
		\hline
		$C_\mu$                  & $0.09$ \\
		\hline
		$\sigma_k$               & $1.0$  \\
		\hline
		$\sigma_\epsilon$        & $1.3$  \\
		\hline
	\end{tabularx}
	\label{tab-coeff-ke}
\end{table}


\bibliographystyle{elsarticle-num}
\bibliography{reference}
%
%
\end{document}